\documentclass[twocolumn,9pt,letterpaper]{revtex4}

\usepackage{hyperref}
\usepackage{amsmath}
\usepackage{amsfonts}
\usepackage{amssymb}
\usepackage{graphicx}
\usepackage{empheq}

\hypersetup{colorlinks=true, urlcolor=blue, linkcolor=blue, citecolor=blue,plainpages=false}

\newcommand{\txtpow}[1]{{\mbox{\scriptsize{#1}}}}

\newcommand{\myeqref}[1]{(\ref{#1})}
\renewcommand{\eqref}[1]{Eq.~(\ref{#1})}
\begin{document}
\title{Dielectric tuning and coupling of whispering gallery modes using an
	anisotropic prism}

\author{Matthew R. Foreman$^{1}$}
\email[]{matthew.foreman@mpl.mpg.de}
\author{Florian Sedlmeir$^{1,2}$}
\author{Harald G. L. Schwefel$^{2}$}
\author{Gerd Leuchs$^{1,3}$}

\affiliation{$^1$Max Planck Institute for the Science of Light,
G\"unther-Scharowsky-Stra{\ss}e 1, 91058 Erlangen, Germany\\
$^2$Department of Physics, University of Otago, Dunedin, New Zealand\\
$^3$Institut f\"ur Optik, Information und Photonik,
Universit\"at Erlangen-N\"urnberg, Staudtstra{\ss}e 7/B2, 91058 Erlangen, Germany}
\date{\today}

\begin{abstract}
Optical whispering gallery mode (WGM) resonators are a powerful and versatile tool used in many branches of science. Fine tuning of the central frequency and line width of individual resonances is however desirable in a number of applications including frequency conversion, optical communications and efficient light-matter coupling. To this end we present a detailed theoretical analysis of dielectric tuning of WGMs supported in
axisymmetric resonators. Using the Bethe-Schwinger equation and
adopting an angular spectrum field representation we study the
resonance shift and mode broadening of high $Q$ WGMs when a planar
dielectric substrate is brought close to the
resonator. Particular focus is given to use of a uniaxial substrate with an
arbitrarily aligned optic axis. Competing red and blue resonance
shifts ($\sim 30$~MHz), deriving from generation of a near field material polarisation and back action from
the radiation continuum respectively, are found. Anomalous resonance
shifts can hence be observed depending on the substrate
material, whereas mode broadening on the order of $\sim 50$~MHz can also be
simply realised. Furthermore, polarisation selective coupling with
extinction ratios of $> 10^4$ can be achieved when the resonator and
substrate are of the same composition and their optic axes are chosen correctly. Double refraction and properties of out-coupled beams are also discussed.
\end{abstract}

\maketitle

\section{Introduction}

Whispering gallery mode (WGM) resonators are a powerful and versatile
tool in modern day optics and have found great employ as novel 
light sources \cite{Sandoghdar1996,Gmachl1998,Spillane2002,Furst2011},
in
spectroscopic studies \cite{Gagliardi2014}, frequency comb generation \cite{Savchenkov2008,Kippenberg2011}, quantum
electrodynamics \cite{Vernooy1998a,Aoki2006},
sensing \cite{Foreman2015a}, nonlinear optics
\cite{Ilchenko2004,Strekalov2016} and optomechanics \cite{Matsko2009,Schliesser2010}. Such extensive usage derives from
the narrow bandwidths, high field strengths and small modal volumes
boasted by WGMs. Primarily, the resonance structure of WGM resonators is dictated by their
geometry and material composition, however, tunability of resonance properties,
such as central frequency and line width, is desirable in
a number of applications. For example, WGM microcavities can be used
as tunable filters or switches, which play a central role in optical signal
processing and classical communication networks
\cite{Monifi2013}. Alternatively, matching the frequency and bandwidth of WGMs to
those of atomic transitions \cite{Schunk2015} enables
efficient light-matter coupling required in quantum communications
\cite{Kimble2008} and information processing
\cite{Matsukevich2004}. Controlled coupling and differential tuning of
WGM resonances can, moreover, greatly improve the efficiency of
polarisation conversion \cite{Smirnov2002} and nonlinear optical
processes such as second harmonic generation or parametric
oscillation \cite{Breunig2016, Sturman2011} and is
therefore important for realisation of novel light sources and
frequency converters.

A number of strategies for tuning WGM resonances based
on: external temperature control \cite{Aoki2006}; electro-optical
\cite{Savchenkov2003} or thermo-optical effects \cite{Armani2004,Ward2010,Lin2015}; and variation of an applied strain
\cite{VonKlitzing2001,Pollinger2009e,Madugani2012} or pressure
\cite{Ioppolo2007,Henze2011}, have previously been reported in the
literature. So-called dielectric
tuning, in which a dielectric substrate is brought into close proximity
to a WGM resonator, has also been proposed and demonstrated as a route to continuous
fine tuning \cite{Schunk2015,Schunk2016}. In this work, we present 
a detailed theoretical formalism describing dielectric frequency
tuning of WGM resonances using planar
substrates, in addition to quantifying substrate induced line width changes. Emphasis is especially placed on the use of arbitrary uniaxial dielectric
planar substrates, which we show can enable differential tuning  and selective coupling of transverse
electric (TE) and transverse magnetic (TM) WGMs \cite{Sedlmeir2016}. 
The structure of this article is therefore as
follows. In Section~\ref{sec:WGMs} we first generalise
the analytic WGM profiles given in \cite{Breunig2013} to account for the open
nature of WGM resonators and detail the properties of WGMs in an isolated
uniaxial axisymmetric resonator required in subsequent
derivations. After establishing a suitable angular
spectrum representation of WGMs in Section~\ref{sec:pert_modes} we
continue to determine the interface induced coupling and resulting
mode distributions in the presence of a uniaxial dielectric prism
(note we shall use the terms prism and substrate interchangeably
throughout this text).
Resonance shifts and mode broadening induced 
by introduction of a planar substrate are subsequently derived by exploiting the
Bethe-Schwinger cavity perturbation formula  in
Section~\ref{sec:shifts}. Critically, far-field
contributions, which can give rise to anomalous radiative shifts 
in open resonators \cite{Ruesink2015}, are incorporated into the
Bethe-Schwinger equation. Section~\ref{sec:results} proceeds to
present a number of illustrative numerical results, before we finally
conclude in Section~\ref{sec:conclusions}.

\section{Whispering gallery modes in open axisymmetric
	resonators}\label{sec:WGMs}

\begin{figure}[!b]
	\begin{center}
		\includegraphics[width=\columnwidth]{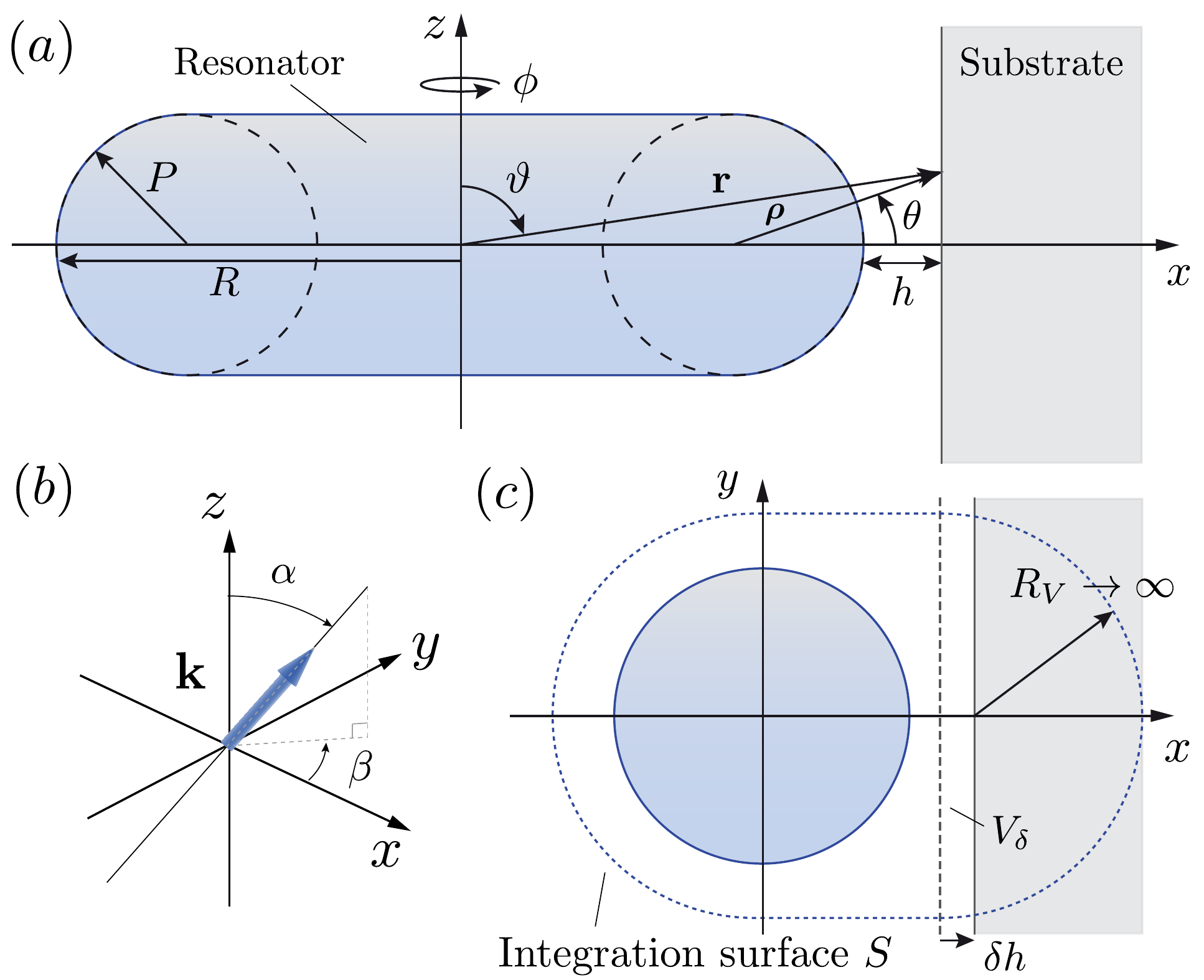}
		\caption{\textbf{WGM resonator geometry}: (a) Definition of
			the coordinate system and geometrical parameters of an
			axisymmetric resonator perturbed by a planar substrate. (b)
			Coordinate system in the far field defining the angular
			dependence of a plane wave component with wavevector
			$\mathbf{k}$ (c) Schematic of integration volume and
			surface used to evaluate the resonance shift and mode
			broadening upon an infinitesimal shift of the dielectric
			substrate away from the resonator.\label{fig:geometry}}
	\end{center}
\end{figure} 
In this section we derive a number of WGM properties that will be required in later sections. We consider a WGM with vacuum wavelength $\lambda = 2\pi/k$ supported in a
uniaxial axisymmetric resonator in air with major radius $R$ and for
which the radius of curvature of the outer surface in the polar
direction (rim radius) is $P$ (see
Figure~\ref{fig:geometry}(a)). We denote the ordinary and extraordinary
refractive index of the resonator by $n_o$ and $n_e$ respectively. Throughout this work we assume that $R \gg
\lambda$ and $P \gg \lambda$ and that the optic axis of the resonator is parallel to the
symmetry axis ($z$-cut). With these assumptions the WGM is well described using
a scalar Helmholtz equation \cite{Breunig2013}. Within the
resonator interior ($\rho \leq P$) the associated electric field
distribution can then be expressed in the form
\begin{align}
\mathbf{E}_{\mathbf{m}}^\nu(\mathbf{r}) \approx
\mathcal{E}_0\exp\!\left[-\frac{\theta^2}{2\Theta_m^2}\right]
\!H_p\!\left[\frac{\theta}{\Theta_m}\right]
\!\mbox{Ai}\left[f^\nu_{\mathbf{m}}(\rho)\right] e^{i m \phi}
\hat{\boldsymbol{\sigma}}_{\txtpow{res}}^\nu ,
\label{eq:WGMmode_int}
\end{align}
where, with reference to Figure~\ref{fig:geometry}(a), $\mathbf{r}=(x,y,z)$ is the position vector relative to the
centre of the resonator, $\rho = |\boldsymbol{\rho}|$ is the distance from the centre of curvature
of the resonator rim, $\theta$ is the toroidal polar angle, $\phi$ is the azimuthal angle, $H_p[z]$ are the Hermite polynomials of degree $p$,  $\mathbf{m} = (m,p,q)$ jointly describes the
azimuthal, polar and radial indices,  $\nu = \mbox{TE}$ or
$\mbox{TM}$ is the polarisation index and $\mathcal{E}_0$ is an arbitrary normalisation constant which is henceforth taken to be unity for simplicity. \eqref{eq:WGMmode_int} was originally
derived assuming a Dirchlet boundary condition at the resonator
surface \cite{Breunig2013}, however, we account for the openness of the
resonator through use of the effective radius $\mathcal{R}_\nu = R +
\Delta_\nu$ where $\Delta_\nu =  \xi/\kappa$ describes the
penetration depth of the WGM field into the host medium, $\kappa = k \sqrt{n^2 -
	1}$, $\xi= 1/n^2=1/n_o^2$ for TM modes and $\xi =1$, $n = n_e$ for
TE modes \cite{Demchenko2013,Ornigotti2014}, such that
\begin{align}
f^\nu_{\mathbf{m}}(\rho) &= ({P} + \Delta_\nu - \rho)/{u_m}-\zeta_q, \\
\Theta_m &= \mathcal{R}^{3/4}_\nu P^{-3/4}  m^{-1/2},\\ 
u_m &= 2^{-1/3} \mathcal{R}_\nu \,m^{-2/3},
\end{align}
where $\zeta_q$ is the $q$-th root of the Airy function
$\mbox{Ai}[-\zeta]=0$. The unit polarisation vectors in \eqref{eq:WGMmode_int} are given by
$\hat{\boldsymbol{\sigma}}_{\txtpow{res}}^{\txtpow{TE}} =-\hat{\boldsymbol{\theta}}$ and
$\hat{\boldsymbol{\sigma}}_{\txtpow{res}}^{\txtpow{TM}} =
\hat{\boldsymbol{\rho}}$, where we have here chosen to neglect the azimuthal polarisation component of the TM mode within the resonator due to its small magnitude relative to the radial component (note that we shall use the caret notation
exclusively to denote unit vectors for which $|\hat{\mathbf{u}}| =
1$). 

Due to the strongly confined nature of WGMs the total energy stored within the resonator
volume, $V$, can be found by integrating over a toroidal volume defined by the local curvature of the resonator as shown in Figure~\ref{fig:geometry}, yielding
\begin{align}
U_{\mathbf{v}}^{\txtpow{res}}
&\approx \frac{1}{2} \epsilon_0n^2\! \int_0^{2\pi} \!\! \int_0^{2\pi}\!\! \int_0^P |\mathbf{E}^\nu_{\mathbf{m}}(\mathbf{r})|^2
\, \rho|R-P + \rho \cos\theta| d\rho d\theta d\phi \label{eq:Umodedef}
\end{align}
where we have also introduced the vector $\mathbf{v} =
(\nu,m,p,q)$ containing all four mode indices. Throughout this work we
shall freely switch between using the combined mode index
$\mathbf{v}$ and separating the polarisation and scalar indices as
$\nu$ and $\mathbf{m}$ depending on whether the polarisation index
must be emphasised. For large $m$ and small $p$ the mode is highly localised near the outer
surface $\rho \approx P$ and at the
equator of the resonator whereby we can make the small angle
approximation $\sin\theta \approx\theta \approx z/\rho$. Combining
Eqs.~\myeqref{eq:WGMmode_int} and \myeqref{eq:Umodedef} hence gives
\begin{align}
U_{\mathbf{v}}^{\txtpow{res}} \!&\approx  \pi \epsilon_0n^2 R\! \int_{-\infty}^{\infty}\!\!\! \exp\!\left[-\frac{z^2}{P^2\Theta_m^2}\right]
\!H_p^2\!\left[\frac{z}{P\Theta_m}\right] dz\nonumber\\
&\hspace{2.5cm}\times  \!\int_0^P\!
\!\mbox{Ai}^2\!\left[f^\nu_{\mathbf{m}}(\rho)\right] d\rho. 
\end{align}
Use of the orthogonality relation of the Hermite functions $\int_{-\infty}^{\infty} \exp[-z^2]H_p[z]H_a[z] dz = \sqrt{\pi} 2^p p! \delta_{ap}$ where $\delta_{ap}$ is the Kronecker delta, then yields
\begin{align}
U_{\mathbf{v}}^{\txtpow{res}} \approx 2^p p! \pi^{3/2} \epsilon_0 u_m  n^2 \Theta_m R P\,  [g_{\mathbf{v}}(0)-g_{\mathbf{v}}(P)] \label{eq:U}
\end{align}
where $g_{\mathbf{v}}(\rho) = f^\nu_{\mathbf{m}}(\rho)
\mbox{Ai}[f^\nu_{\mathbf{m}}(\rho)]^2 -
\mbox{Ai}'[f^\nu_{\mathbf{m}}(\rho)]^2$ and the prime notation denotes
differentiation with respect to the argument. For large $m$ we note
that $g_{\mathbf{v}}(0) \approx 0$ and $g_{\mathbf{v}}(P) \approx - \mbox{Ai}'[-\zeta_q]^2$.

Exterior to the resonator ($\rho \geq P$) the WGM exhibits an evanescent decay \cite{Demchenko2013} viz.
\begin{align}
\hspace{-2.8mm}\,\mathbf{E}_{\mathbf{m}}^\nu(\mathbf{r})  \approx A_\nu
\exp\left[-\frac{\theta^2}{2\Theta_m^2}\right]
H_p\left[\frac{\theta}{\Theta_m}\right] e^{-\kappa (\rho-P)} e^{i m \phi}\hat{\boldsymbol{\sigma}}_{\txtpow{sur}}^\nu\label{eq:WGMmode_ext}
\end{align}
where $A_\nu$ is a normalisation constant required to match the fields at the resonator surface, $\hat{\boldsymbol{\sigma}}_{\txtpow{sur}}^{\txtpow{TE}} = -\hat{\boldsymbol{\theta}}$, $\hat{\boldsymbol{\sigma}}_{\txtpow{sur}}^{\txtpow{TM}} = i a_{\txtpow{sur}}\hat{\boldsymbol{\phi}} -b_{\txtpow{sur}}\hat{\boldsymbol{\rho}} $ and $a_{\txtpow{sur}}/b_{\txtpow{sur}} \approx  (1-n_o^{-2})^{1/2}$ as follows from application of the Fresnel coefficients for light undergoing total internal reflection at a glancing angle to the resonator surface \cite{Junge2013}. 
From applying the usual Maxwell boundary conditions it follows that
$A_{\txtpow{TE}} = \mbox{Ai}[\Delta_{\txtpow{TE}}/u_m -
\zeta_q]$. Care must be taken when matching TM modes however since in
this case the magnitude of the azimuthal field component becomes
$({n_o^4 -n_o^2})^{1/2}$ times greater than that of the radial
component at the (interior) resonator surface and thus must also  be
included. With this in mind it can be shown that $A_{\txtpow{TM}}
=(2n_o^4-n_o^2)^{1/2} \mbox{Ai}[\Delta_{\txtpow{TM}}/u_m - \zeta_q]
$. Moreover, given \eqref{eq:WGMmode_ext} it follows that the
relative contribution to the mode energy from the field outside the
resonator is 
\begin{align}
\frac{U_{\mathbf{v}}^{\txtpow{sur}}}{U_{\mathbf{v}}^{\txtpow{res}}}
\approx \frac{A_\nu^2\int_P^\infty
	\exp[-2\kappa(\rho-P)]d\rho}{n^2\int_0^P
	\mbox{Ai}^2[f_{\mathbf{m}}^\nu(\rho)] d\rho} \approx \frac{A_\nu^2}{2
	n^2\kappa u_m \mbox{Ai}'[-\zeta_q]^2} \label{eq:Uext}
\end{align} 
from which it quickly follows that $U_{\mathbf{v}}^{\txtpow{res}} \gg
U_{\mathbf{v}}^{\txtpow{sur}}$ such that the majority of the mode energy is
seen to lie within the resonator. \eqref{eq:Uext} holds within
the approximation that the field outside the resonator decays
exponential, however, similar expressions can also be derived
through use of the complete mode profile \cite{Gagliardi2014,Foreman2014a}. Finally, we note that although expressions for the WGM mode profile and energy given in this section are approximate in nature, they have been found to agree well with rigorous finite element calculations \cite{Oxborrow2007}.

\section{Mode distributions in the presence of an anisotropic
	interface}\label{sec:pert_modes}

Interaction of a WGM with a dielectric perturbation, gives rise to a
redistribution of the mode profile. This phenomenon has been well studied for the
case of perturbing dielectric and plasmonic nanoparticles, in addition to consideration of layered structures \cite{Teraoka2006a,Teraoka2007a,Foreman2015,Foreman2013c}. Changes in the properties of morphological dependent resonances of a dielectric sphere near a conducting plane have also been considered using a multipolar expansion \cite{Johnson1994,Johnson1996a}. Here we consider
the mode distribution, $\mathbf{E}_{\mathbf{v}}(\mathbf{r};h)$,
resulting from interaction of a WGM with a semi-infinite dielectric
substrate  whose interface is placed at a 
distance $h$ from the surface of a $z$-cut birefringent resonator and with its normal directed
along $\hat{\mathbf{x}}$ as depicted in
Figure~\ref{fig:geometry}(a). The mode profiles defined in
Eqs.~\myeqref{eq:WGMmode_int} and
\myeqref{eq:WGMmode_ext} represent the limiting case $h\rightarrow \infty$.
Moreover, we consider a uniaxial dielectric substrate
with arbitrary optic axis $\hat{\mathbf{c}}$ and (extra-)ordinary
refractive index ($n_e^{\txtpow{sub}}$) $n_o^{\txtpow{sub}}$. To account for potential back coupling into the
resonator from reflection of the WGM from the interface, we first
consider surface
dressing of the WGM amplitude. With this in hand, we then proceed
to derive the mode distribution in both the resonator interior and
exterior, in addition to that in the infinite half space of the anisotropic
dielectric, using an angular spectrum approach.

\subsection{Surface dressed scattering amplitudes}\label{sec:surfacedressing}
To describe excitation of WGMs in an isolated resonator structure, we can
consider an arbitrary incident field, which we represent as a
superposition of modes viz.
\begin{align}
\mathbf{E}_i(\mathbf{r}) = \sum_{\mathbf{v}} a_{\mathbf{v}} \mathbf{V}_{\mathbf{v}} (\mathbf{r}).
\end{align}
Assuming the modes form a complete orthogonal basis over the surface, $A$, of
the resonator it follows that 
\begin{align}
a_{\mathbf{v}} = \frac{\iint_A \mathbf{V}_{\mathbf{v}}^*(\mathbf{r}) \cdot
	\mathbf{E}_i(\mathbf{r})  dA}{ \iint_A \mathbf{V}_{\mathbf{v}}^*(\mathbf{r})\cdot
	\mathbf{V}_{\mathbf{v}}(\mathbf{r})  dA }. \label{eq:coeff_proj}
\end{align}
The incident
field gives rise to a scattered field, which can again be represented as a
superposition of modes as
\begin{align}
\mathbf{E}_s(\mathbf{r};h \rightarrow \infty)  = \sum_{\mathbf{v}} b_{\mathbf{v}} \mathbf{W}_{\mathbf{v}}(\mathbf{r}) ,
\end{align}
where the additional parametric dependence on $h$ is introduced for later convenience.
Note that we consider different modes $\mathbf{V}_{\mathbf{v}}$ and
$\mathbf{W}_{\mathbf{v}}$ due to differing physical requirements at infinity. Specifically, the scattered modes must satisfy the
Sommerfeld radiation condition, whereas the incident modes have a zero
net energy flow through a closed surface and physically cannot possess
a singularity in the volume of interest. 
Through application of the Maxwell boundary conditions at the surface of the
resonator, the incident and scattered mode coefficients can be related
through the matrix equation $\mathbf{b} = \mathbb{N}\mathbf{a}$, where
$\mathbb{N}$ is the scattering matrix and $\mathbf{a} =
[a_{\mathbf{v}}]$ ($\mathbf{b} =
[b_{\mathbf{v}}]$) is a vector formed by stacking all the incident (scattered) mode
coefficients. For a spherical resonator, the modes $\mathbf{V}_{\mathbf{v}}$ and
$\mathbf{W}_{\mathbf{v}}$ would correspond to the vector multipole
modes with a radial dependence described by the spherical Bessel
functions and Hankel functions (of the first kind) respectively. Accordingly, $\mathbb{N}$ would be a
diagonal matrix with non-zero elements given by the usual Mie
scattering amplitudes \cite{Bohren1983a}. Modal properties of
resonances in isolated optical resonators, such as the associated
resonance frequencies and line widths, can then be determined through analysis
of the poles of the scattering matrix $\mathbb{N}$
\cite{McVoy1967}. The field in the interior of the resonator can
similarly be written
\begin{align}
\mathbf{E}_r(\mathbf{r};\infty) = \sum_{\mathbf{v}} f_{\mathbf{v}} \mathbf{U}_{\mathbf{v}}(\mathbf{r})
\end{align}
where the internal mode coefficients can also be related to the
illumination coefficients through 
$\mathbf{f} = \mathbb{Z} \mathbf{a}$.

Introduction of a dielectric inhomogeneity in
the resonator surroundings, e.g. a dielectric interface or a perturbing
nanoparticle, produces two additional contributions to the total field
exterior to the resonator.
Firstly, the incident field $\mathbf{E}_i$
is reflected from the inhomogeneity giving rise to a field
$\mathbf{E}_{ir}$.  Similarly, the field scattered from the resonator is reflected
giving rise to a field $\mathbf{E}_{sr}$.  Each contribution can also
be decomposed  according to \footnote{Note that the choice of modes
	to use in this decomposition, in general, depends on the position
	$\mathbf{r}$ at which the field is considered. In this work
	we assume that the dielectric inhomogeneity is located exterior to
	the resonator and we consider the field close to the resonator
	such that \eqref{eq:ref_field_decomp} holds.}
\begin{align}
\mathbf{E}_{ir} (\mathbf{r};h)&= \sum_{\mathbf{v}} c'_{\mathbf{v}}
\mathbf{V}_{\mathbf{v}}(\mathbf{r}) \nonumber
\\
\quad\mathbf{E}_{sr}(\mathbf{r};h) &= \sum_{\mathbf{v}}
d'_{\mathbf{v}} \mathbf{V}_{\mathbf{v}}(\mathbf{r}). \label{eq:ref_field_decomp}
\end{align}
The total field incident upon the
resonator is that formed from the superposition of the incident 
and the reflected fields, i.e. $\mathbf{E}_i^{\txtpow{eff}} =
\mathbf{E}_i + \mathbf{E}_{ir}+ \mathbf{E}_{sr}$, such that the scattered field is 
\begin{align}
\mathbf{E}_{s} (\mathbf{r};h)= \sum_{\mathbf{v}} b'_{\mathbf{v}} \mathbf{W}_{\mathbf{v}}(\mathbf{r}),
\end{align}
where $\mathbf{b}' = \mathbb{N} (\mathbf{a} + \mathbf{c}'+\mathbf{d}')$
and $\mathbf{b}'$, $\mathbf{c}'$ and $\mathbf{d}'$ are perturbed
coefficient vectors defined
analogously to above. Noting further, that the reflected incident field originates from the
incident field, we may write $\mathbf{c}' =
\mathbb{J}\mathbf{a}$ and similarly for the reflected scattered
field we have $\mathbf{d}' = \mathbb{K} \mathbf{b}'$. Solving for the
scattered coefficients $\mathbf{b}'$ in terms of the illumination coefficients $\mathbf{a}$, thus
yields
\begin{align}
\mathbf{b}' = [\mathbb{I} - \mathbb{N}\mathbb{K}]^{-1}
\mathbb{N}[\mathbb{I} + \mathbb{J}] \mathbf{a} \triangleq \mathbb{N}_{\txtpow{eff}}
\mathbf{a} \label{eq:dressed_N}
\end{align}
where $\mathbb{I}$ is the identity matrix  and $\mathbb{N}_{\txtpow{eff}}$ is an effective, or dressed,
scattering matrix. Through an analogous argument it also follows that
\begin{align}
\mathbf{f}' = \mathbb{Z} ( \mathbb{I} + \mathbb{J} + \mathbb{K}
\mathbb{N}_{\txtpow{eff}}) \mathbf{a} \triangleq
\mathbb{Z}_{\txtpow{eff}} \mathbf{a}. \label{eq:dressed_Z}
\end{align}
Excitation of WGMs is typically achieved through
evanescent coupling using a prism or waveguide structure
\cite{Gorodetsky1999}, such that the excitation field is non-negligible over only a
small extent of the resonator surface. Practically, dielectric tuning
is also achieved using an independent dielectric substrate \cite{Schunk2016,Sedlmeir2016}, such that the
excitation field does not contribute to
the total field at the substrate interface. Accordingly we can safely
neglect the contribution of $\mathbf{E}_{ir}$ for our purposes
(i.e. we assume $\mathbb{J} = \mathbb{O}$, where $\mathbb{O}$ is the
null matrix). We briefly note that the effective scattering
coefficient derived above is equivalent to that which would be found
by considering multiple reflections of the WGM from the dielectric substrate and summing over all
reflected orders. 

\subsection{Prism induced mode coupling \label{sec:modecoupling}}

Notably, evaluation of the effective scattering and transmission
matrices described by \eqref{eq:dressed_N} and \eqref{eq:dressed_Z} requires
determination of the complex coupling coefficients,
$K_{\mathbf{u},\mathbf{v}}$, which comprise $\mathbb{K}$, which we consider in in detail in this section. Our derivation consists
of three steps: firstly we determine the angular spectrum of the WGM
with mode index $\mathbf{v} = (\nu,m,p,q)$ at
the dielectric interface; secondly, each constituent plane wave
component is reflected from the interface by means of generalised
Fresnel reflection coefficients; before finally, calculating the
mode overlap between the reflected field with the WGM of order
$\mathbf{u} = (\mu,l,a,b )=(\mu,\mathbf{l})$ (defined analogously to
$\mathbf{v}$).

The first step of our derivation requires the angular spectrum
$\widetilde{\mathbf{E}}_{\mathbf{m}}^{\nu}(\mathbf{k};h)$ of a given WGM at the
dielectric interface located at $x_0 = R+h$. Restricting our attention to large $m$ WGMs we
can legitimately make a small angle approximation, and with reference
to Figure~\ref{fig:geometry}(a), it follows that $\hat{\boldsymbol{r}}
\approx \hat{\boldsymbol{\rho}}$ and $\hat{\boldsymbol{\theta}} \approx
-\hat{\boldsymbol{\vartheta}}$. The symmetry of the TE and TM WGMs
therefore implies the angular spectrum of the field is of the form $\widetilde{\mathbf{E}}_{\mathbf{m}}^{\txtpow{TE}}(\mathbf{k};h) \approx \widetilde{E}_{\mathbf{m}}^{\txtpow{TE}} (\mathbf{k};h) \hat{\boldsymbol{\alpha}}$ and $\widetilde{\mathbf{E}}_{\mathbf{m}}^{\txtpow{TM}}(\mathbf{k};h) \approx \widetilde{E}_{\mathbf{m}}^{\txtpow{TM}} (\mathbf{k};h) \hat{\boldsymbol{\beta}}$, where $\mathbf{k} =
k[\sin\alpha\cos\beta,\sin\alpha\sin\beta,\cos\alpha]$, and, $\alpha$ and
$\beta$ are the (complex) polar and azimuthal angles in $\mathbf{k}$
space (analogous to $\vartheta$ and $\phi$) as depicted in
Figure~\ref{fig:geometry}(b). We note that the radial polarisation
component is absent in the far field so as to ensure 
transversality of the field. The scalar amplitudes, evaluated in a given $x$ plane, are
given by
\begin{align}
\widetilde{E}_{\mathbf{m}}^\nu(\mathbf{k};x-R) = \frac{1}{4\pi^2}\iint_{-\infty}^{\infty}{E}_{\mathbf{m}}^\nu(x,y,z)
e^{-i(k_y y + k_z z )} dydz \label{eq:FTangspec}
\end{align}
where 
${E}_{\mathbf{m}}^\nu(x,y,z)$ is the complex amplitude of
$\mathbf{E}_{\mathbf{m}}^\nu(x,y,z)$ and we parameterise the
amplitudes in terms of the distance $x-R$ of the $x$ plane from the
resonator rim for later convenience. At the dielectric
interface we have $x=x_0 = (R-P + \rho\cos\theta)\cos\phi$, which for small $\theta$ and
$\phi$ gives $\rho \approx P + h + (P+h)\theta^2/2 +
(R+h)\phi^2/2$. Together with \eqref{eq:WGMmode_ext} we thus find
\begin{align}
E_{\mathbf{m}}^\nu(x_0,y,z) & \approx A_\nu H_p\left[\frac{z}{P\,\Theta_m}\right] \exp\left[-\kappa h + i m \frac{y}{R} \right]\nonumber\\
&\times \exp\left[-\frac{1}{2} \left\{ \frac{y^2}{\Delta y^2}+
\frac{z ^2}{\Delta z^2}+\frac{z^2}{P^2\Theta_m^2}\right\} \right],\label{eq:Es_interface}
\end{align}
where $\Delta y^2 = R/\kappa$, $\Delta z^2 = P/\kappa$ and we have also used the fact that $R \gg h$ and $P \gg h$. \eqref{eq:Es_interface} represents the product of the WGM mode with a Gaussian coupling window, such that use of the convolution theorem yields
\begin{align}
\widetilde{{E}}_{\mathbf{m}}^\nu(\mathbf{k};h)  &\approx \widetilde{A}_\nu \,  \exp\left[- \frac{ 1}{2} \!\left\{
(k_y - k_{yr})^2 \Delta y^2  +
k_z^2  \delta z^2 \right\}\right]  e^{-\kappa h}\nonumber\\
&\quad\times \!\left[ \sum_{q = 0}^{p/2} \frac{p! \, (-i)^{p-2q} }{q! (p-2q)!}
\frac{(-2)^q}{(1+\sigma^2 )^q } H_{p-2q}\!
\left[\frac{  k_z \delta z^2 }{P\Theta_m} \right]\! \right]\label{eq:angspec}
\end{align}
where $\widetilde{A}_\nu = A_\nu  \delta z \Delta y/(2\pi) $, $\delta
z^{-2}  = \Delta z^{-2} + (P\Theta_m)^{-2}$ and $\sigma = \Delta z /(P
\Theta_m)$. From \eqref{eq:angspec} we see that as a consequence of the small
angle approximation the angular spectrum is
separable in $k_y$ and $k_z$. Moreover, in the $k_y$ direction the
angular spectrum is centred on $k_{yr} =  m / R$ (i.e. the
propagation constant of the associated WGM) and has a width of $1/
\Delta y \ll k$, whereas in the $k_z$ direction the spectrum is centred around $k_z =
0$ and has a width $\sim 1/\delta z \ll k$. 

\begin{figure}[!t]
	\begin{center}
		\includegraphics[width=\columnwidth]{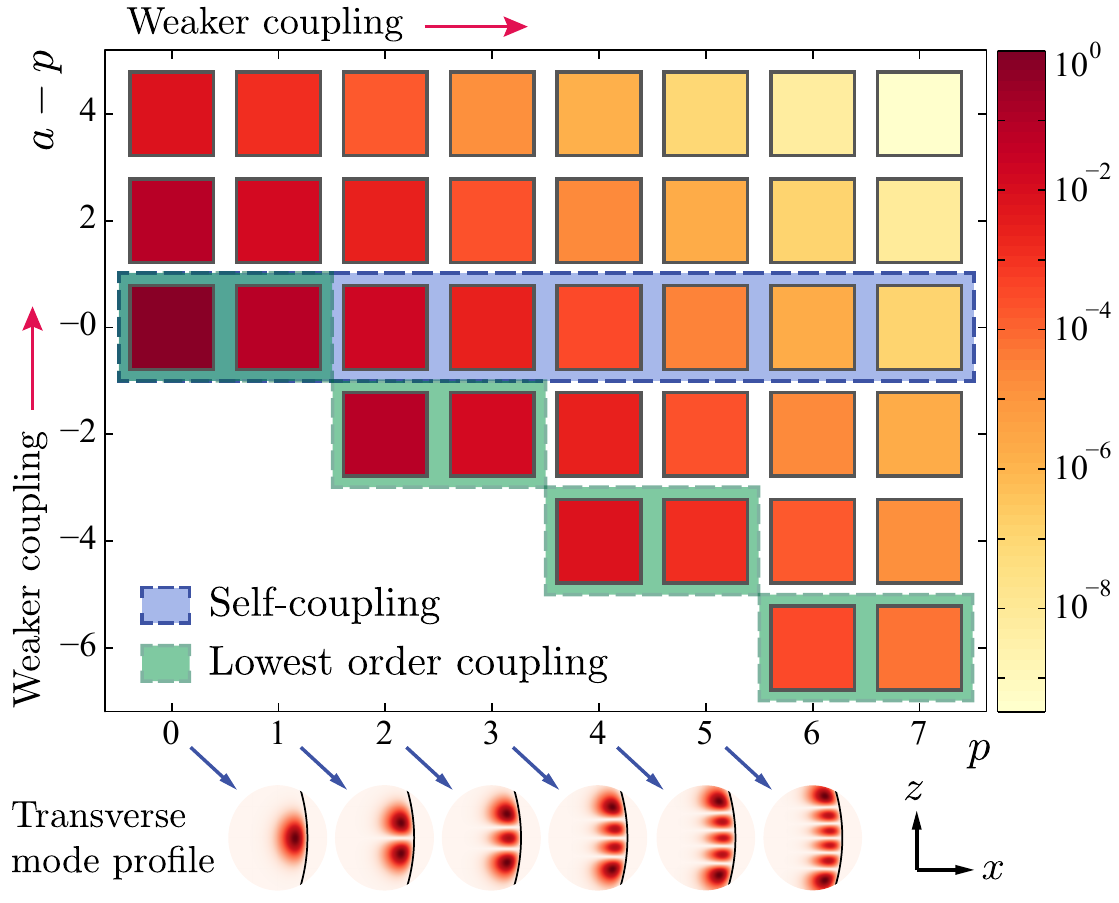}
		\caption{\textbf{Interface induced coupling strengths}: Heat map of
			the coupling coefficients between a WGM of polar order $p$
			to a WGM of polar order $a$, as given by $|I_{ap}^{(2)}|  /
			[2^{a+p+1} p! a!] $. A value of $\tau = 1.2$ was
			assumed. Mode profiles of differing polar orders
			are shown in the lower inset from which different odd-even
			symmetry classes are apparent. Strongest coupling of a
			given WGM order $p$ is to the lowest order mode of the
			same symmetry as highlighted by the green boxes, however,
			coupling to modes of a different symmetry class is not possible. \label{fig:Iap}}
	\end{center}
\end{figure}

Reflection and transmission of each angular component can be described
through use of generalised Fresnel reflection and transmission
coefficients, a derivation of which can be found in
Appendix~\ref{app:Fresnel}. We note, however, that the transmission
coefficients act on the $\hat{\mathbf{s}} = (\mathbf{k}\times \hat{\mathbf{x}})/|\mathbf{k}\times \hat{\mathbf{x}}|$ and
$\hat{\mathbf{p}} = (\mathbf{k}\times \hat{\mathbf{s}})/|\mathbf{k}\times \hat{\mathbf{s}}|$ field components, i.e. those perpendicular and
parallel to the plane of incidence. It is therefore necessary to express the
$\hat{\boldsymbol{\alpha}}$ and $\hat{\boldsymbol{\beta}}$ field
components in terms of the $\hat{\mathbf{s}}$ and
$\hat{\mathbf{p}}$ basis. Upon making the usual small angle
approximation we find $
\widetilde{\mathbf{E}}_{\mathbf{m}}^{\txtpow{TE}}(\mathbf{k};h) \approx
\widetilde{E}^{\txtpow{TE}}_{\mathbf{m}}(\mathbf{k};h) \, \hat{\mathbf{s}}$ and $\widetilde{\mathbf{E}}_{\mathbf{m}}^{\txtpow{TM}}(\mathbf{k};h) \approx
\widetilde{{E}}^{\txtpow{TM}}_{\mathbf{m}} (\mathbf{k};h)  \,
\hat{\mathbf{p}}$. The reflected angular spectrum in each case,
denoted $\widetilde{\mathbf{E}}^{\nu}_{\mathbf{m},r}(\mathbf{k}_r;h)$, is
hence
\begin{align}
\widetilde{\mathbf{E}}^{\txtpow{TE}}_{\mathbf{m},r}(\mathbf{k}_r;h) &=
\widetilde{E}^{\txtpow{TE}}_{\mathbf{m}}(\mathbf{k};h) [ r_{ss} \hat{\mathbf{s}}_r +
r_{sp}   \hat{\mathbf{p}}_r ]\\
\widetilde{\mathbf{E}}^{\txtpow{TM}}_{\mathbf{m},r}(\mathbf{k}_r;h) &=
\widetilde{E}^{\txtpow{TM}}_{\mathbf{m}}(\mathbf{k};h) [r_{ps}
\hat{\mathbf{s}}_r +
r_{pp}  \hat{\mathbf{p}}_r],
\end{align}
where $\mathbf{k}_r = [-k_x,k_y,k_z]$ is the reflected wavevector, $\hat{\mathbf{s}}_r =
\hat{\mathbf{s}}$, 
$\hat{\mathbf{p}}_r = [\mathbf{k}_r\times \hat{ \mathbf{s}}
]/|\mathbf{k}_r\times \hat{\mathbf{s}}|$ and the
reflection coefficients, $r_{ij}$, are defined in Appendix~\ref{app:Fresnel}. Accordingly the reflected field at a general
position $\mathbf{r}$ exterior to the resonator is given by
\begin{align}
\mathbf{E}_{\mathbf{m},r}^{\txtpow{TE}}(\mathbf{r};h) &=
\,  \mathcal{E}_{\mathbf{m}}^{\txtpow{TE,TE}}(\mathbf{r};h) \,\,\hat{\boldsymbol{\sigma}}_{\txtpow{ref}}^{\txtpow{TE}} +
\mathcal{E}_{\mathbf{m}}^{\txtpow{TE,TM}}(\mathbf{r};h) \,\hat{\boldsymbol{\sigma}}_{\txtpow{ref}}^{\txtpow{TM}}\label{eq:Eref}\\
\mathbf{E}_{\mathbf{m},r}^{\txtpow{TM}}(\mathbf{r};h) &=
\mathcal{E}_{\mathbf{m}}^{\txtpow{TM,TE}}(\mathbf{r};h)  \,\hat{\boldsymbol{\sigma}}_{\txtpow{ref}}^{\txtpow{TE}} + \mathcal{E}_{\mathbf{m}}^{\txtpow{TM,TM}}(\mathbf{r};h)
\,\hat{\boldsymbol{\sigma}}_{\txtpow{ref}}^{\txtpow{TM}} \label{eq:Eref2}
\end{align}
as
follows from reciprocity and where
$\hat{\boldsymbol{\sigma}}_{\txtpow{ref}}^{\txtpow{TE}} =
- \hat{\boldsymbol{\theta}}_r$, $\hat{\boldsymbol{\sigma}}_{\txtpow{ref}}^{\txtpow{TM}} =
[ i a_{\txtpow{sur}} \hat{\boldsymbol{\phi}}_r -b_{\txtpow{sur}}
\hat{\boldsymbol{\rho}}_r]$, $ \hat{\boldsymbol{\rho}}_r =
\hat{\boldsymbol{\rho}}(\theta,\pi-\phi)$ and similarly for
$\hat{\boldsymbol{\theta}}_r$ and $\hat{\boldsymbol{\phi}}_r$. The amplitude factors are given by
\begin{align}
\mathcal{E}_{\mathbf{m}}^{\txtpow{TE,TE}} (\mathbf{r};h) &=
\iint_{-\infty}^{\infty} r_{ss}
\widetilde{E}^{\txtpow{TE}}_{\mathbf{m}}(\mathbf{k};h) \exp[i
\mathbf{k}_r\cdot  \Delta\mathbf{r}] dk_y dk_z \label{eq:Eint1}\\
\mathcal{E}_{\mathbf{m}}^{\txtpow{TE,TM}} (\mathbf{r};h) &=
\iint_{-\infty}^{\infty} r_{sp}
\widetilde{E}^{\txtpow{TE}}_{\mathbf{m}}(\mathbf{k};h) \exp[i
\mathbf{k}_r\cdot   \Delta\mathbf{r}] dk_y dk_z \label{eq:Eint2}\\
\mathcal{E}_{\mathbf{m}}^{\txtpow{TM,TE}} (\mathbf{r};h) &=
\iint_{-\infty}^{\infty} r_{ps}
\widetilde{E}^{\txtpow{TM}}_{\mathbf{m}}(\mathbf{k};h) \exp[i
\mathbf{k}_r\cdot  \Delta \mathbf{r}] dk_y dk_z \label{eq:Eint3}\\
\mathcal{E}_{\mathbf{m}}^{\txtpow{TM,TM}} (\mathbf{r};h) &=
\iint_{-\infty}^{\infty} r_{pp}
\widetilde{E}^{\txtpow{TM}}_{\mathbf{m}}(\mathbf{k};h) \exp[i
\mathbf{k}_r\cdot  \Delta \mathbf{r}] dk_y dk_z. \label{eq:Eint4}
\end{align}
where $ \Delta\mathbf{r} = (x-x_0,y,z) = \mathbf{r}-x_0\hat{\mathbf{x}}$ is the position vector relative to the
plane of the prism interface.
Determination of the reflected field over the resonator surface requires
evaluation of these integrals for $\mathbf{r}\in A$. To do so we make a number of further
approximations. We once more note that $k_y \approx m/R$, $k_z \approx
0$, $\Delta k_y \ll k$ and $\Delta k_z \ll k$, such that $-i k_x x
\approx \kappa x$. Furthermore given the small range of $k_y$ and
$k_z$, we assert that the variation of the reflection coefficients is
small, such that $r_{ij}(k_y,k_z) \approx r_{ij}(m/R,0)
\triangleq \bar{r}_{ij}$
for $\{i,j\} = \{s,p\}$ whereby the reflection coefficient can be 
factored outside of the
integral. With these approximations
Eqs.~\myeqref{eq:Eint1}--\myeqref{eq:Eint4} essentially reduce to the
2D inverse Fourier transform of the angular spectra
$\widetilde{E}^{\nu}_{\mathbf{m}}(\mathbf{k};h)$ with some additional Gaussian
factors due to the additional evanescent decay of the mode. 
We thus find on the resonator surface, i.e. $(x,y,z) \in A$, that
\begin{align}
\mathcal{E}_{\mathbf{m}}^{\nu,\mu}(x,y,z;h) &\approx A_{\nu} \bar{r}_{ij}
H_p\left[\frac{z}{P\Theta_m}\right] \exp\left[-2\kappa h + i m
\frac{y}{R}\right] \nonumber\\
&\quad\quad\times \exp\left[- \left\{ \frac{y^2}{\Delta y^2}+
\frac{z ^2}{\Delta z^2}+\frac{z^2}{2P^2\Theta_m^2}\right\} \right],\label{eq:Er_interface}
\end{align}
where $i=s$ ($p$) for $\nu = \mbox{TE}$ (TM) and similarly for
$j$. With this association understood we henceforth adopt the shorthand notation
$\bar{r}_{\nu\mu} = \bar{r}_{ij}$.
The coupling constant can then be evaluated viz. (c.f. \eqref{eq:coeff_proj}) 
\begin{align}
K_{\mathbf{u},\mathbf{v}} =\frac{ \iint_A {\mathbf{E}_{\mathbf{l}}^{\mu *}}(\mathbf{r})
	\cdot \mathbf{E}^{\nu}_{\mathbf{m},r}(\mathbf{r};h) 
}{\iint_A
{\mathbf{E}_{\mathbf{l}}^{\mu *}}(\mathbf{r})\cdot
{\mathbf{E}_{\mathbf{l}}^{\mu}}(\mathbf{r}) } dA\triangleq \frac{I^{(1)}_{\mathbf{u},\mathbf{v}}}{I_{\mathbf{u}}^{(0)}}.
\end{align}
We first evaluate the integral $I_{\mathbf{u}}^{(0)}$ which follows easily from
the orthogonality of the Hermite functions as
\begin{align}
I_{\mathbf{u}}^{(0)}&\approx 2^{a+1}\pi^{3/2}A_\mu^2 \,a! \,RP\,\Theta_l.
\end{align}
Turning attention to evaluation of $I^{(1)}_{\mathbf{u},\mathbf{v}}$,
we first consider the polarisation dependence. Within our small
angle approximation it can be shown that
$\hat{\boldsymbol{\sigma}}_{\txtpow{sur}}^{\nu,*}\cdot
\hat{\boldsymbol{\sigma}}_{\txtpow{ref}}^{\mu} \approx
1$ for $\nu =\mu=
\mbox{TE}$ and $\approx k^2 / (2k_{yr}^2 - k^2) \approx (2 n_o^2-1)^{-1}$ for $\nu =\mu=
\mbox{TM}$ and zero otherwise. Moreover, the narrow windowing function
described by the $\exp[-y^2/\Delta y^2]$ factor means that we need only
perform the integration over a small region of the resonator surface
whereby $dA \approx dy dz$. Use of Eqs.~\myeqref{eq:Es_interface} and
\myeqref{eq:Er_interface} then allows us to express $I^{(1)}_{\mathbf{u},\mathbf{v}}$ in the form
\begin{align}
I^{(1)}_{\mathbf{u},\mathbf{v}}&\approx 
\,\pi^{1/2} A_\nu A_\mu  \bar{r}_{\nu\mu} \Delta y P \Theta_m
I^{(2)}_{ap} \hat{\boldsymbol{\sigma}}_{\txtpow{sur}}^{\mu,*}\cdot
\hat{\boldsymbol{\sigma}}_{\txtpow{ref}}^{\mu}  \nonumber\\
&\hspace{2cm}\times \exp[-2 \kappa h]\exp\left[{-\frac{(m-l)^2}{2\Delta m^2}} \right]\label{eq:Iuv1_partial}
\end{align}
where $\Delta m^2 \approx 2m^{-1}\sqrt{1-n^{-2}}$. For large $m$ we
have $\Delta m \ll 1$, such that the second exponential in
\eqref{eq:Iuv1_partial} can be safely replaced by the Kronecker delta
$\delta_{l,m}$ whereby $\Theta_l = \Theta_m$. The
remaining integral term, $  I^{(2)}_{ap} $, in \eqref{eq:Iuv1_partial}
can then be written as
\begin{align}
I^{(2)}_{ap} &= \int_{-\infty}^{\infty}
H_a\left[w\right]  H_{p}\left[w\right] \exp\left[-2\tau^2w^2\right] d w
\end{align}
with $\sqrt{2}\tau = P\Theta_m / \delta z$. It is evident that $I^{(2)}_{ap}$  is identically zero when $a$ and $p$ are of opposite parity,
i.e. polar modes of differing symmetry do not couple as would be
expected. When $a+p$ is even, however, $I_{ap}^{(2)}$ can be evaluated
analytically \cite{Erdelyi1954a} yielding 
\begin{align}
I^{(2)}_{ap} &=2^{\frac{a+p-1}{2}} s^{-a-p-1} (1-2
\tau^2)^{\frac{a+p}{2}}
\,  \Gamma\left[\frac{a+p+1}{2}\right] \nonumber\\
&\quad\quad\quad\times  {}_2F_1\left[-a,-p;\frac{1-a-p}{2};\frac{\tau^2}{2\tau^2 - 1}\right],
\end{align}
where $\Gamma[n]$ and ${}_2F_1[a,b;c;z]$ are the Gamma and Gauss
hypergeometric functions respectively. Hence, we
ultimately arrive at the desired coupling coefficients
\begin{align}
K_{\mathbf{u},\mathbf{v}} = \delta_{l,m}
\frac{\bar{r}_{\nu\mu} \Delta y 
	I^{(2)}_{ap}  }{2^{a+1} \pi a! R }  \,e^{-2\kappa h}\times
\left\{\begin{array}{ll} 
\!\!1 &\mu = \mbox{TE} \\
\!\!(2n_o^2 - 1)^{-1} &\mu = \mbox{TM.} 
\end{array}\right. \label{eq:K}
\end{align}
A heat map depicting the interface induced
coupling strength between modes of different polar orders ($a$ and
$p$) as described by \eqref{eq:K} is shown in Figure~\ref{fig:Iap}. Note that for ease
of comparison we have also
included a further normalisation factor of $2^pp!$ in the data shown in
Figure~\ref{fig:Iap}, which arises from
the scaling of the initial mode energy (\eqref{eq:Umodedef}).
Inspection of \eqref{eq:K} and Figure~\ref{fig:Iap} reveals that a WGM of order $p$
couples most strongly to the lowest order WGM of the same symmetry upon reflection,
i.e.  the fundamental $a=0$ mode for even $p$ or the $a=1$ mode for
odd $p$. This is a direct consequence of the finite width coupling
window. We reiterate that
coupling between modes of differing symmetry is not
possible. Moreover, as follows from the conservation of angular
momentum, coupling between modes with different azimuthal indices
($l$ and $m$) is also forbidden. \eqref{eq:K}, however, does not
forbid coupling between modes of differing polarisation, albeit,
within our small angle approximation, such polarisation mixing stems
from the anisotropy of the substrate and would be absent for an
isotropic prism.

\subsection{Mode distributions}
Having described the reflection of a WGM by a dielectric substrate, we now determine the complete perturbed mode distributions. WGMs with differing mode indices are typically spectrally distinct in
most resonators (with the exception of a perfectly spherical
resonator for which the polar modes are degenerate) as is reflected in
the amplitude of the scattering coefficients, $\eta_{\mathbf{v}}$ and $\zeta_{\mathbf{v}}$, contained in $\mathbb{N}$
and $\mathbb{Z}$ respectively. Consequently, assuming $R\neq P$, we can make a single
mode approximation whereby \eqref{eq:dressed_N}, simplifies to
\begin{align}
b_{\mathbf{v}}' \approx \frac{\eta_{\mathbf{v}}}{1 - \eta_{\mathbf{v}} K_{\mathbf{v},\mathbf{v}}}
a_{\mathbf{v}}\triangleq {\eta_{\mathbf{v},\txtpow{eff}}} \, a_{\mathbf{v}}.\label{eq:dressed_b}
\end{align}
Close to resonance and initially neglecting possible material absorption
in the resonator, the scattering coefficient can be approximated by
the Breit–Wigner line shape \cite{Johnson1993} viz. 
\begin{align}
\eta_{\mathbf{v}}(\omega) =  -\frac{ \gamma_{\mathbf{v},\txtpow{rad}}/2}{ \gamma_{\mathbf{v},\txtpow{rad}}/2-i(\omega - \omega_{\mathbf{v}})
} \label{eq:Lorentz_approx}
\end{align}
where $\omega_{\mathbf{v}}$ and $\gamma_{\mathbf{v},\txtpow{rad}}$ denote the resonance
frequency and radiative line width respectively.  The $-1$ pre-factor  follows from imposing field continuity at the resonator surface.  As shown in
Appendix~\ref{app:Lorentzian} material absorption in the
resonator reduces the magnitude of the peak scattering amplitude such that
\begin{align}
\eta_{\mathbf{v}}(\omega) = - \frac{\gamma_{\mathbf{v},\txtpow{rad}}/2}{(\gamma_{\mathbf{v},\txtpow{rad}}+\gamma_{\mathbf{v},\txtpow{abs}})/2 -i(\omega - \omega_{\mathbf{v}})} \label{eq:Lorentz_approx2}
\end{align}
where $\gamma_{\mathbf{v},\txtpow{abs}} $ is the absorptive line width
\cite{Gorodetsky1996}. Similarly \eqref{eq:dressed_Z} simplifies
to
\begin{align}
f_{\mathbf{v}}' \approx \frac{\zeta_{\mathbf{v}}}{1-\eta_{\mathbf{v}} K_{\mathbf{v},\mathbf{v}}} a_{\mathbf{v} } \triangleq
\zeta_{\mathbf{v},\txtpow{eff}} a_{\mathbf{v}}. \label{eq:dressed_f}
\end{align}
Combination of Eqs.~\myeqref{eq:dressed_b},
\myeqref{eq:Lorentz_approx2}  and \myeqref{eq:dressed_f} shows that on
resonance the scattered and internal mode coefficients $b_{\mathbf{v}}'$ and $f_{\mathbf{v}}'$
are reduced by a factor of $\mathcal{A}(h) = (1-\eta_{\mathbf{v}} K_{{\mathbf{v}},{\mathbf{v}}})^{-1}  \approx
(\gamma_{{\mathbf{v}},\txtpow{rad}} + \gamma_{{\mathbf{v}},\txtpow{abs}})   /
(\gamma_{{\mathbf{v}},\txtpow{rad}} + \gamma_{{\mathbf{v}},\txtpow{abs}} +
\gamma_{{\mathbf{v}},\txtpow{rad}} K_{{\mathbf{v}},{\mathbf{v}}} )$ relative to the
unperturbed ($h\rightarrow \infty$) case. For absorption limited resonators for which
$\gamma_{{\mathbf{v}},\txtpow{abs}}  \gg \gamma_{{\mathbf{v}},\txtpow{rad}}$ we find
$\mathcal{A}(h)\approx 1$ and hence $b_{\mathbf{v}}' \approx
b_{\mathbf{v}}$ and $f_{\mathbf{v}}' \approx f_{\mathbf{v}}$.  Within the resonant mode approximation, it therefore immediately
follows that the WGM field distribution within the resonator ($\rho
< P$) in the
presence of a dielectric interface is given simply by
$\mathbf{E}_{\mathbf{v}}(\mathbf{r} ;h ) = \mathcal{A}(h)
\mathbf{E}_{\mathbf{v}} (\mathbf{r};\infty)$, where
$\mathbf{E}_{\mathbf{v}} (\mathbf{r};\infty)$ is given by
\eqref{eq:WGMmode_int}. 
Recalling results from above we also find the field exterior to the resonator ($\rho
\geq P$, $x < x_0$) is given by
$\mathbf{E}_{\mathbf{v}}(\mathbf{r} ;h ) = \mathcal{A}(h) [
\mathbf{E}_{\mathbf{v}} (\mathbf{r};\infty) + \mathbf{E}_{\mathbf{v},r}(\mathbf{r};h)]$, where now 
$\mathbf{E}_{\mathbf{v}} (\mathbf{r};\infty)$ is given by
\eqref{eq:WGMmode_ext} and
$\mathbf{E}_{\mathbf{v},r}(\mathbf{r};h)$ follows from
Eqs.~\myeqref{eq:Eref}-\myeqref{eq:Eint4}. Following the same line of
arguments used to derive \eqref{eq:Er_interface} we can, however, write
\begin{align}
\mathbf{E}_{\mathbf{m},r}^\nu(\mathbf{r};h) 
&\approx \sum_{\mu\in\{\txtpow{TE,TM}\}} \bar{r}_{\nu\mu}\exp[-2 \kappa h] 
E_{\mathbf{v}}(- x,y,z;\infty)
\hat{\boldsymbol{\sigma}}_{\txtpow{ref}}^\mu .  \label{eq:Er_approx}
\end{align}

The field transmitted  into the volume of the anisotropic substrate $x\geq x_0$ can be found similarly to the
reflected field using the generalised
Fresnel transmission coefficients (Appendix~\ref{app:Fresnel}). From
the preceding analysis in Section~\ref{sec:modecoupling} we note that for an initially unperturbed WGM
of order $\mathbf{v}$, the component of the perturbed field incident
on the prism interface is given by $\mathcal{A}(h)
\mathbf{E}_{\mathbf{m}}^{\nu}(x_0,y,z;\infty)$, with a corresponding angular spectrum
$\mathcal{A}(h)
\widetilde{\mathbf{E}}_{\mathbf{m}}^\nu(\mathbf{k};h)$. Upon
transmission, each constituent plane wave generates an ordinary and
extraordinary wave in the prism, with associated wavevectors
$\mathbf{k}_{o}$  and $\mathbf{k}_{e}$, such that the transmitted
field takes the form
\begin{align}
\mathbf{E}^{\nu}_{\mathbf{m}}(\mathbf{r};h) = \mathcal{A}(h) \sum_{j\in\{o,e\}}
\iint_{-\infty}^{\infty} t_{ij}  \widetilde{E}^{\nu}_{\mathbf{m}}(\mathbf{k};h) &
\hat{\mathbf{j}} \, e^{i \mathbf{k}_j\cdot\Delta\mathbf{r}} 
dk_y dk_z,\label{eq:Et_NF} 
\end{align}
where the subscript $i = s,p$ for $\nu = \mbox{TE, TM}$ respectively, $t_{ij}$ ($j
\in \{o,e\}$) are the generalised
Fresnel transmission coefficients and $\hat{\mathbf{j}} = \hat{\mathbf{o}}$ and $\hat{\mathbf{e}}$ are the unit
polarisation vectors for the ordinary and extraordinary waves
respectively. The explicit dependence of $t_{ij}$, $\mathbf{k}_{o}$  and
$\mathbf{k}_{e}$ on the incident wavevector $\mathbf{k}$ is given in
Appendix~\ref{app:Fresnel}. To simplify \eqref{eq:Et_NF} further we note that exponential terms vary
rapidly with $k_y$ and $k_z$. In comparison the Fresnel coefficients and
polarisation terms vary weakly within the limited range of $k_y$ and
$k_z$ over which $\widetilde{\mathbf{E}}^{\nu}_{\mathbf{m}}$ is
non-negligible. As such we make the approximations
$\hat{\mathbf{o}} \approx (\overline{\mathbf{k}}_o \times \hat{\mathbf{c}} )\, /
|\overline{\mathbf{k}}_o \times \hat{\mathbf{c}} |$
and $\hat{\mathbf{e}} \approx  \tensor{\epsilon}{}^{-1} (\overline{\mathbf{k}}_e \times [\overline{\mathbf{k}}_e \times \hat{\mathbf{c}}  ]) /
|\tensor{\epsilon}{}^{-1} (\overline{\mathbf{k}}_e \times [\overline{\mathbf{k}}_e \times \hat{\mathbf{c}} ])|$,
where $\overline{\mathbf{k}}_o$ and $\overline{\mathbf{k}}_e$ are the
central ordinary and extraordinary wavevectors found from
Eqs.~\myeqref{eq:kox}--\myeqref{eq:wdef} with $k_y = k_{yr}$ and
$k_z=0$. 
We hence obtain
\begin{align}
&\mathbf{E}^{\nu}_{\mathbf{m}}(\mathbf{r};h) = \nonumber \\
&\,\, \mathcal{A}(h)\sum_{j\in\{o,e\}}\bar{t}_{ij} \,
\hat{\mathbf{j}} 
\iint_{-\infty}^{\infty} \widetilde{E}^{\nu}_{\mathbf{m}}(\mathbf{k};h) 
e^{i {k}_{j,x}\Delta x} e^{i(k_y
	y + k_z z)}   dk_y dk_z
\label{eq:Et_NF_TETM_app}
\end{align}
where  $\Delta x = x-x_0$ and 
$\bar{t}_{ij}$ is defined analogously to
$\bar{r}_{ij}$ above. Once more taking advantage of the small angular
spread in $k_y$ and $k_z$ we expand the $k_{j,x}$ exponent around $k_y
= k_{yr}$ and $k_z = 0$ which to leading order yields
\begin{align}
k_{o,x} &\approx (n_o^2 k^2-k_y k_{yr}) / (n_o^2 k^2 -
k_{yr}^2)^{1/2} \triangleq \chi_{o}^{(0)} + \chi_{o}^{(1)} k_y\label{eq:kox_exp}\\
k_{e,x} &\approx \frac{2 k^2}{\overline{D}^{1/2}} +
\frac{k_y}{k_{yr}}\left[\frac{\overline{v} + \overline{D}^{1/2}}{2u} - \frac{2
	k^2}{\overline{D}^{1/2}}   \right] \triangleq \chi_{e}^{(0)} + \chi_{e}^{(1)} k_y\label{eq:kex_exp}
\end{align}
where $\overline{D} = \overline{v}^2 - 4 u \overline{w}$, $\overline{v} = v(k_{yr},0)$ and $\overline{w} =
w(k_{yr},0)$ follow from Eqs.~\myeqref{eq:udef}--\myeqref{eq:wdef} in Appendix~\ref{app:Fresnel}. The
constant terms in Eqs.~\myeqref{eq:kox_exp} and \myeqref{eq:kex_exp}
can be factored out of the integrals of \eqref{eq:Et_NF_TETM_app} such
that the integrals are now of the form of a simple Fourier transform
with respect to the variables $k_y$ and $k_z$. Accordingly, the terms
linear in $k_y$ can be simply accounted for by recalling the shift
theorem, whereby \eqref{eq:Et_NF_TETM_app} becomes
\begin{align}
&\mathbf{E}^{\nu}_{\mathbf{m}}(\mathbf{r};h) =\nonumber\\ &\quad\mathcal{A}(h)\sum_{j\in\{o,e\}}
\bar{t}_{ij} \,
\hat{\mathbf{j}}\, e^{i \chi_j^{(0)}\Delta x}
{E}^{\nu}_{\mathbf{m}}\left(x_0,y+
\chi_j^{(1)} \Delta x,z;\infty\right)  \label{eq:Et_NF_TETM_app2}  
\end{align}
where $\chi_j^{(0)}$ and $\chi_j^{(1)}$ are defined by Eqs.~\myeqref{eq:kox_exp} and \myeqref{eq:kex_exp}.
We note that near the critical angle, the variation of the transmission
coefficients $t_{ij}$ can vary strongly with $k_y$ such that the
approximations leading to \eqref{eq:Et_NF_TETM_app2}  are not
valid. In this case the full integral expressions of \eqref{eq:Et_NF}
must be used to account for Fresnel filtering effects
\cite{Rex2002,Tureci2002}. Similar restrictions also apply to
calculation of the reflected field and mode coupling as discussed in Section~\ref{sec:modecoupling}.

\section{Prism induced resonance perturbations}\label{sec:shifts}
Shifts in the resonance frequency of modes in a closed cavity induced
by local dielectric perturbations can be described by the
Bethe-Schwinger equation \cite{Waldron1960}. Moreover, it has recently been
shown that this formula can also be used to account for mode
broadening and radiative shifts in open resonators if the
far-field components are incorporated \cite{Ruesink2015}. In
Appendix~\ref{app:BSderiv} we briefly present a derivation of the
Bethe-Schwinger equation for open cavities in the presence of anisotropic dielectric
perturbations. Although ultimately it is our goal to determine the
total resonance shift and line width broadening induced by the presence
of a uniaxial substrate relative to the case in which no substrate
is present, the perturbed and unperturbed modes required to evaluate
the Bethe-Schwinger equation (\eqref{eq:BetheSchwinger}) can not be
taken as $\mathbf{E}_{\mathbf{v}}(\mathbf{r};\infty)$
and  $\mathbf{E}_{\mathbf{v}}(\mathbf{r};h)$ given above. This
can intuitively be seen, since the second term in
\eqref{eq:BetheSchwinger} can be associated with radiative losses in
the far field, however, this term evaluates to zero if the unperturbed
mode $\mathbf{E}_{\mathbf{v}}(\mathbf{r};\infty)$ is chosen because in
the far field $|\mathbf{E}_{\mathbf{v}}(\mathbf{r};\infty)|
\rightarrow 0$. Physically such a scenario is incorrect since
radiative losses give rise to a mode broadening which is not accounted
for. Instead, we consider the change of the WGM frequency
and lifetime assuming that the dielectric is initially located at a
distance $h$ from the resonator and is then displaced by an
infinitesimally small distance, $\delta h$ away from the
resonator. Accordingly, the Bethe-Schwinger equation takes the form
\begin{align}
\delta \overline{\omega}_{\mathbf{v}} =
\overline{\omega}_{\mathbf{v}}(h + \delta h) -
\overline{\omega}_{\mathbf{v}}(h) = -\left[\overline{\omega}{
	N_{\mathbf{v}}^{(1)}}+ i { N^{(2)}_{\mathbf{v}}} \right]  \big/
N_{\mathbf{v}}^{(0)}, \label{eq:BSmain}
\end{align}
where we have defined the integral terms (assuming $\delta
\mu = \mathbb{O}$)
\begin{align}
N_{\mathbf{v}}^{{(0)}}&=\int_V  \Big\{
\mathbf{E}^{*}_{\mathbf{v}}(\mathbf{r};h) \, \tensor{\epsilon} \, \mathbf{E}_{\mathbf{v}}(\mathbf{r};h+\delta h
) \nonumber\\&\hspace{2.5cm} 
+\mathbf{H}^{*}_{\mathbf{v}}(\mathbf{r};h) \, \tensor{\mu} \, \mathbf{H}_{\mathbf{v}}(\mathbf{r};h+\delta h
)\Big\}dV,\label{eq:N0def}\end{align}
\begin{align}
N^{(1)}_{\mathbf{v}}&= \int_{V_\delta}                           
\mathbf{E}^{*}_{\mathbf{v}}(\mathbf{r};h)  \, \delta
\tensor{\epsilon} \, 
\mathbf{E}_{\mathbf{v}}(\mathbf{r};h+\delta h
) \,dV,\label{eq:N1def}\\
N^{(2)}_{\mathbf{v}}&=\oint_S \Big\{\delta\mathbf{E}^{*}_{\mathbf{v}}(\mathbf{r}) \times  \mathbf{H}_{\mathbf{v}}(\mathbf{r};h+\delta h
) \nonumber\\&\hspace{2.5cm}+
\mathbf{E}_{\mathbf{v}}^*(\mathbf{r};h+\delta h) \times \delta\mathbf{H}_{\mathbf{v}}(\mathbf{r}) \Big\} \cdot d\mathbf{S} \label{eq:N2def}.
\end{align}
and $\delta \mathbf{E}_{\mathbf{v}}(\mathbf{r}) =
\mathbf{E}_{\mathbf{v}}(\mathbf{r};h+\delta h) -
\mathbf{E}_{\mathbf{v}}(\mathbf{r};h) $ (and similarly for $\delta\mathbf{H}$).
$V_{\delta}$ defines the volume  $h\leq x-R \leq h+\delta h$ for which 
$\delta\tensor{\epsilon} = \epsilon_0 \mathbb{I} -
\tensor{\epsilon}{}^{\txtpow{sub}}$ is non zero. The volume $V$ and the associated surface $S$ over
which the integrals in $N_{\mathbf{v}}^{(0)}$, $N_{\mathbf{v}}^{(1)}$
and $N_{\mathbf{v}}^{(2)}$ are taken are shown in
Figure~\ref{fig:geometry}(c). The total change in the complex resonance frequency
induced by the presence of the substrate can then be found by noting 
\begin{align}
\frac{d\overline{\omega}}{dh} = \lim\limits_{\delta h \rightarrow 0}
\frac{\delta \overline{\omega}}{\delta h}\label{eq:BS_diff}
\end{align}
such that 
\begin{align}
\Delta \overline{\omega}_{\mathbf{v}} =
\overline{\omega}_{\mathbf{v}}(h) -
\overline{\omega}_{\mathbf{v}}(\infty) = \int_{\infty}^h
\frac{d\overline{\omega}}{dh'}  dh'. \label{eq:BS_final_int}
\end{align}
Each of the $N_{\mathbf{v}}$ terms defined in Eqs.~\myeqref{eq:N0def}--\myeqref{eq:N2def} will now be considered in turn.

\paragraph{Evaluation of $N_{\mathbf{v}}^{(0)}$:}
We initially consider evaluation of the volume integral
$N_{\mathbf{v}}^{(0)}$. Since we consider an infinitesimal shift of
the dielectric substrate it follows that the volume of the perturbation
in $\tensor{\epsilon}(\mathbf{r})$ is much smaller than the total
volume considered. Accordingly we can approximate
$N_{\mathbf{v}}^{(0)}$ as \cite{Ruesink2015}
\begin{align}
N_{\mathbf{v}}^{(0)}&=\int_V  \Big\{
\mathbf{E}^{*}_{\mathbf{v}}(\mathbf{r};h) \, \tensor{\epsilon} \, \mathbf{E}_{\mathbf{v}}(\mathbf{r};h)
+\mathbf{H}^{*}_{\mathbf{v}}(\mathbf{r};h) \, \tensor{\mu} \,
\mathbf{H}_{\mathbf{v}}(\mathbf{r};h)\Big\}dV\label{eq:N0defapprox}
\end{align}
which is four times the total energy of the WGM in the presence
of a dielectric substrate placed at a distance $h$ from the resonator
surface. We further approximate the
latter, by the surface dressed mode energy of  an unperturbed WGM, whereby
\begin{align}
N_{\mathbf{v}}^{(0)}&= 4|\mathcal{A}(h)|^2{U_{\mathbf{v}}^{\txtpow{res}}} \label{eq:denom_res}
\end{align}
where $U_{\mathbf{v}}^{\txtpow{res}}$ is the mode energy defined by Eqs.~\myeqref{eq:Umodedef} and \myeqref{eq:U}.

\paragraph{Evaluation of $N_{\mathbf{v}}^{(1)}$: }
To evaluate $N_{\mathbf{v}}^{(1)}$
we note that within the domain of integration, $V_{\delta}$, the mode distribution before the interface
is displaced is that which is transmitted into the
substrate, i.e. that defined by \eqref{eq:Et_NF}.
After the interface is shifted the field distribution is given by $
\mathcal{A}(h+\delta h) [
\mathbf{E}_{\mathbf{v}} (\mathbf{r}) +
\mathbf{E}_{\mathbf{v},r}(\mathbf{r};h+\delta h)]$. From
Eqs.~\myeqref{eq:Eref} and \myeqref{eq:Et_NF} it therefore follows
that
\begin{widetext}
	\begin{align}
	N_{\mathbf{v}}^{(1)} &= \mathcal{A}^*(h) \mathcal{A}(h+\delta h)\int_{V_{\delta}}
	\iint_{-\infty}^{\infty}\iint_{-\infty}^{\infty}  \widetilde{E}^*_{\mathbf{v}}(\mathbf{k};h)
	\widetilde{E}_{\mathbf{v}}(\mathbf{k}';h+\delta h)
	\nonumber\\
	&\times\left[\sum_{j \in \{
		o,e\}} t_{ij} \exp[i \mathbf{k}_j \cdot\Delta \mathbf{r}]  \hat{\mathbf{j}}\right]^*
	\delta\tensor{\epsilon} 
	\left[\sum_{j' \in \{ s,p\}} \delta_{ij'}\exp[i
	\mathbf{k}'\cdot \Delta\mathbf{r}] e^{-i k_x' \delta h} \hat{\mathbf{j}}'+ r_{ij'} \exp[i \mathbf{k}_r' \cdot
	\Delta\mathbf{r}]e^{i k_x' \delta h}  \hat{\mathbf{j}}'_r\right] dk_ydk_z dk_y'dk_z' dV
	\end{align}
	where $i = s,p$ for $\nu = \mbox{TE, TM}$ as previously discussed.
	Rearranging the order of integration, the integrals over $y$ and $z$ can be evaluated immediately by
	noting that $\int_{-\infty}^{\infty} \exp[i(k_y'-k_y)y]dy = 2\pi
	\delta(k_y'-k_y)$ and similarly for $k_z$. Consequently integration
	over $k_y'$ and $k_z'$ can also be simply performed yielding
	\begin{align}
	N_{\mathbf{v}}^{(1)} 
	= 4\pi^2|\mathcal{A}(h)|^2 \sum_{j \in\{
		o,e\}}&\sum_{j' \in\{ s,p\}} 
	\int_{x_0}^{x_0+\delta h}
	\!\! \iint_{-\infty}^{\infty}
	|\widetilde{E}_{\mathbf{v}}(\mathbf{k};h)|^2 
	t_{ij}^*  \nonumber\\
&\quad\times 	\left[\delta_{ij'} e^{-(i {k}^*_{j,x} +\kappa) \Delta x } \hat{\mathbf{j}}^*
	\delta\tensor{\epsilon} \hat{\mathbf{j}}'+ r_{ij'}
	e^{-(i {k}_{j,x}^* - \kappa) \Delta x } e^{-2\kappa \delta h}  \,\hat{\mathbf{j}}^*
	\delta\tensor{\epsilon} \hat{\mathbf{j}}'_r\right]
	dk_ydk_z dx +
	O(\delta h)
	\end{align}
\end{widetext}
\vspace{-0.75cm}{\flushleft{where}} we have expanded $\mathcal{A}(h+\delta h) \approx \mathcal{A}(h)
+ \delta h \, d\mathcal{A}/dh+\ldots$, used \eqref{eq:angspec} to write
$\widetilde{E}_{\mathbf{v}}(\mathbf{k}';h+\delta h) =
\widetilde{E}_{\mathbf{v}}(\mathbf{k}';h) \exp[- \kappa\delta h]$ and
made the substitution $-i k_x \approx \kappa$. The integration over
$x$ can also be performed analytically such that we obtain
\begin{align}
N_{\mathbf{v}}^{(1)} &= 4\pi^2 |\mathcal{A}(h)|^2 \sum_{j \in \{
	o,e\}}\sum_{j' \in\{ s,p\}} \iint_{-\infty}^{\infty} t_{ij}^*
|\widetilde{E}_{\mathbf{v}}(\mathbf{k};h)|^2 \nonumber\\
&\times\left[\delta_{ij'} \mathcal{X}_j^+  \hat{\mathbf{j}}^*
\delta\tensor{\epsilon} \hat{\mathbf{j}'} + r_{ij} \mathcal{X}_j^-
e^{-2\kappa \delta h}   \,\hat{\mathbf{j}}^*
\delta\tensor{\epsilon} \hat{\mathbf{j}'_r} \right] dk_y dk_z\nonumber \\
&\quad\quad\quad\quad +
O(\delta h)\label{eq:N1full}
\end{align}
where
\begin{align}
\mathcal{X}_j^\pm &= \int_{x_0}^{x_0+\delta h} \exp\left[-(ik_{j,x}^* \pm
\kappa)(x-x_0)\right]dx\\
&= \frac{1- \exp[-(ik_{j,x}^* \pm\kappa)\delta h]}{i k_{j,x}^* \pm \kappa}.
\end{align}

\paragraph{Evaluation of $N_{\mathbf{v}}^{(2)}$:}
We begin by separating $N_{\mathbf{v}}^{(2)}$ into two distinct integrals: one over the
hemisphere in the half-space $x\geq x_0$ (in the prism) and another
over a hemisphere in the half-space $x< x_0$
(in the host medium),  which we denote by $
N_{\mathbf{v}}^{(2>)} + N_{\mathbf{v}}^{(2<)}$ respectively. In the
dielectric substrate we can use \eqref{eq:Et_NF} and write
\begin{align}
\delta\mathbf{E}_{\mathbf{v}}(\mathbf{r}) &=
\mathcal{A}(h) \sum_{j \in \{o,e\}}
\iint_{-\infty}^{\infty} \Big[ t_{ij} \widetilde{E}_{\mathbf{v}}(\mathbf{k};h)
e^{i
	\mathbf{k}_j \cdot \Delta\mathbf{r} } \, \hat{\mathbf{j}}
\nonumber\\
&\quad\times    \left\{\exp[-(ik_{j,x} + \kappa)\delta h] - 1\right\}  \Big] dk_y dk_z +
O(\delta h).\label{eq:deltaE}
\end{align}
We note, however, that we have chosen our
integration surface such that $R_V
\rightarrow \infty$ (see Appendix~\ref{app:BSderiv}). Consequently, through application of the method
of stationary phase \cite{Mandel1995a}, \eqref{eq:deltaE} becomes
\begin{align}
&\lim\limits_{|\Delta\mathbf{r}| \rightarrow\infty}\delta\mathbf{E}_{\mathbf{v}}(\mathbf{r}) =
-2\pi i\mathcal{A}(h) \nonumber\\
&\quad\times\sum_{j\in\{o,e\}} 
t_{ij}  k_{j,x}\widetilde{E}_{\mathbf{v}}(\mathbf{k};h)
\left\{e^{-(ik_{j,x} + \kappa)\delta h} - 1\right\}  \frac{e^{i k_j \Delta
		r}}{\Delta r} \,  \hat{\mathbf{j}}\nonumber\\
&\quad\quad\quad\quad+
O(\delta h),\label{eq:deltaE_ff}
\end{align}
where $\Delta r = |\Delta \mathbf{r}|$ and $k_j =
|\mathbf{k}_j|$. Further noting
${\omega}\mu_0\widetilde{\mathbf{H}}_{\mathbf{v}}
(\mathbf{k};h)= \mathbf{k} \times
\widetilde{\mathbf{E}}_{\mathbf{v}} (\mathbf{k};h)$ and that in the
far field $d\mathbf{S} = \Delta \hat{\mathbf{r}}  dS =
\hat{\mathbf{k}}_j dS$ 
we have for a fixed direction
\begin{widetext}
	\begin{align}
&\lim\limits_{|\Delta\mathbf{r}|
		\rightarrow\infty}[\delta\mathbf{E}_{\mathbf{v}}(\mathbf{r})
	\times \mathbf{H}^*_{\mathbf{v}}(\mathbf{r}) ] \cdot\hat{\mathbf{k}}_j = \frac{4\pi^2}{\omega\mu_0}
	\frac{|\mathcal{A}(h)|^2 }{\Delta r^2} \nonumber \\ &\quad\quad\quad\quad\times\sum_{j\in\{o,e\}}\sum_{j'\in\{o,e\}}
	t_{ij}t_{ij'}^* k_{j,x}k_{j',x}
	|\widetilde{\mathbf{E}}_{\mathbf{v}} (\mathbf{k};h)|^2
	\left\{\exp[-(i k_{j,x} + \kappa)\delta h ]-1\right\} \,
	\hat{\mathbf{j}} \times ({\mathbf{k}}_{j'} \times
	\hat{\mathbf{j}}') \cdot{\hat{\mathbf{k}}_{j}} +O(\delta h)
	\end{align}
\end{widetext}
{\flushleft{where}} we have also restricted to propagating waves
($\mbox{Im}[k_{j,x}] = 0$) since evanescent components do not
contribute to the far field. Using the standard identity $\mathbf{a}\times(\mathbf{b}\times
\mathbf{c}) = \mathbf{b}(\mathbf{a}\cdot\mathbf{c}) -
\mathbf{c}(\mathbf{a}\cdot\mathbf{b})$ yields
\begin{align}
\hat{\mathbf{j}} \times ({\mathbf{k}}_{j'} \times
\hat{\mathbf{j}}') \cdot{\hat{\mathbf{k}}_{j}} = \delta_{jj'} k_j
[1-(\hat{\mathbf{j}}\cdot \hat{\mathbf{k}}_j)^2],
\end{align}
where $\hat{\mathbf{o}}\cdot \hat{\mathbf{k}}_j =0$ but
$\hat{\mathbf{e}}\cdot \hat{\mathbf{k}}_j \neq 0$ due to the
anisotropic nature of the substrate. It can easily be shown that $\lim_{|\Delta\mathbf{r}|
	\rightarrow\infty}[\delta\mathbf{E}_{\mathbf{v}}(\mathbf{r})
\times \mathbf{H}^*_{\mathbf{v}}(\mathbf{r}) ]
\cdot\hat{\mathbf{k}}_j =  \lim_{|\Delta\mathbf{r}|
	\rightarrow\infty}[\mathbf{E}_{\mathbf{v}}^*(\mathbf{r})
\times \delta\mathbf{H}_{\mathbf{v}}(\mathbf{r}) ]
\cdot\hat{\mathbf{k}}_j$, and that the surface element is given by $dS = \Delta
r^2 \sin\vartheta d\vartheta d \phi = \Delta r^2 dk_y dk_z / (k_j
k_{j,x})$, whereby it follows that 
\begin{align}
\hspace{-0.3cm}& N_{\mathbf{v}}^{(2 >)} = \frac{8\pi^2}{\omega \mu} |\mathcal{A}(h)|^2
\sum_{j\in\{o,e\}} \iint\limits_{\txtpow{~~~Im}[k_{j,x}] = 0} \Big[k_{j,x} |t_{ij}|^2
|\widetilde{\mathbf{E}}_{\mathbf{v}} (\mathbf{k};h)|^2 \nonumber\\
&\quad\,\times  \left\{e^{-(i k_{j,x} + \kappa)\delta h }-1\right\}
\left\{1-(\hat{\mathbf{j}}\cdot \hat{\mathbf{k}}_j)^2\right\} \Big] dk_y dk_z +
O(\delta h). \label{eq:N2full}
\end{align}
In the half space $x<x_0$ a similar analysis can be performed as that
presented for $x\geq x_0$. Critically, for large resonators the
majority of the plane wave components in the medium surrounding the
resonator are
evanescent in nature and do not contribute in the far field. From
the law of reflection it also follows that the reflected plane wave components of the WGM are
also evanescent in nature, such that we can conclude that $N_{\mathbf{v}}^{(2<)} \approx 0$.

\paragraph{Complex resonance shifts:} Upon substituting
Eqs.~\myeqref{eq:denom_res}, \myeqref{eq:N1full} and
\myeqref{eq:N2full} into \eqref{eq:BSmain} we note that the
$|\mathcal{A}(h)|^2$ factors cancel from the leading terms in both the
numerator and denominator. Upon taking the limit in
\eqref{eq:BS_diff} we thus find
\begin{align}
\frac{d \overline{\omega}}{dh} = -\frac{1}{U_{\mathbf{v}}^{\txtpow{res}}}\left[\overline{\omega}  \mathcal{N}_{\mathbf{v}}^{(1)}(h) + i  \mathcal{N}_{\mathbf{v}}^{(2)}(h)\right] 
\end{align}
where
\begin{align}
\mathcal{N}_{\mathbf{v}}^{(1)}(h) &= \pi^2 \sum_{j \in\{
	o,e\}}\sum_{j' \in\{ s,p\}} \iint_{-\infty}^{\infty}t_{ij}^* 
|\widetilde{E}_{\mathbf{v}}(\mathbf{k};h)|^2 \nonumber\\
&\hspace{2.2cm}\times\left[\delta_{ij'} \hat{\mathbf{j}}^*
\delta\tensor{\epsilon} \hat{\mathbf{j}'} + r_{ij} \,\hat{\mathbf{j}}^*
\delta\tensor{\epsilon} \hat{\mathbf{j}'_r} \right] dk_y dk_z  \label{eq:mathcalN1}\\
\mathcal{N}_{\mathbf{v}}^{(2)}(h) &=- \frac{2\pi^2}{\omega \mu}
\sum_{j 
	\in\{ o,e\}} \iint\limits_{\txtpow{~~~Im}[k_{j,x}] = 0} k_{j,x} (i
k_{j,x} + \kappa)|t_{ij}|^2
\nonumber\\
&\hspace{1.4cm}\times |\widetilde{\mathbf{E}}_{\mathbf{v}} (\mathbf{k};h)|^2  \left[1-(\hat{\mathbf{j}}\cdot \hat{\mathbf{k}}_j)^2\right] \, dk_y dk_z . \label{eq:mathcalN2}
\end{align}
Finally, noting $ |\widetilde{E}_{\mathbf{v}}(\mathbf{k};h)|^2 =
|\widetilde{E}_{\mathbf{v}}(\mathbf{k};0)|^2 \exp[-2\kappa h]$, we
can 
evaluate \eqref{eq:BS_final_int} to obtain
\begin{align}
\Delta \overline{\omega}_{\mathbf{v}}(h)=
\frac{\exp[- 2 \kappa h]}{2\kappa U_{\mathbf{v}}^{\txtpow{res}}}\left[\overline{\omega}
\mathcal{N}_{\mathbf{v}}^{(1)}(0) + i
\mathcal{N}_{\mathbf{v}}^{(2)}(0)\right],\label{eq:final_shift}
\end{align}
from which the shift in the (real) resonance frequency and the change
in the line width follow as $\Delta \omega_{\mathbf{v}} = \mbox{Re}[\Delta
\overline{\omega}_{\mathbf{v}}]$ and $\Delta \gamma_{\mathbf{v}} = -2
\mbox{Im}[\Delta \overline{\omega}_{\mathbf{v}}]$
respectively. \eqref{eq:final_shift} constitutes the key result of
this article and will form the basis of the remaining analysis.

\section{Numerical results and discussion}\label{sec:results}

To study the resonance perturbations induced by a (uniaxial) dielectric substrate we now apply \eqref{eq:final_shift} to a number of
scenarios. In all calculations we consider a lithium niobate (LiNbO$_3$)
resonator supporting WGMs at $\lambda\approx 1550$~nm, whereby
the ordinary and extraordinary refractive indices are given by $n_o =
2.213$ and $n_e = 2.138$ respectively \cite{Zelmon1997}. Furthermore, the
major and minor radii of the resonator are taken to be $R = 2.1$~mm
and $P = R/9$ respectively and the optic axis of the resonator is
assumed to be parallel to its axis of rotation (i.e. $z$-cut). Accordingly, the fundamental ($p =
0, q= 1$) TE and TM
resonances at $\lambda \approx 1550$~nm, as found using the dispersion
relation given in \cite{Breunig2013}, are
of order $m = 18152$ and $18790$ respectively. The associated
amplitude decay lengths  are $\kappa^{-1}\approx 130$~nm and
$125$~nm.

\subsection{Anomalous radiative shifts}
\begin{figure*}[t!]
	\begin{center}
		\includegraphics[width=\textwidth]{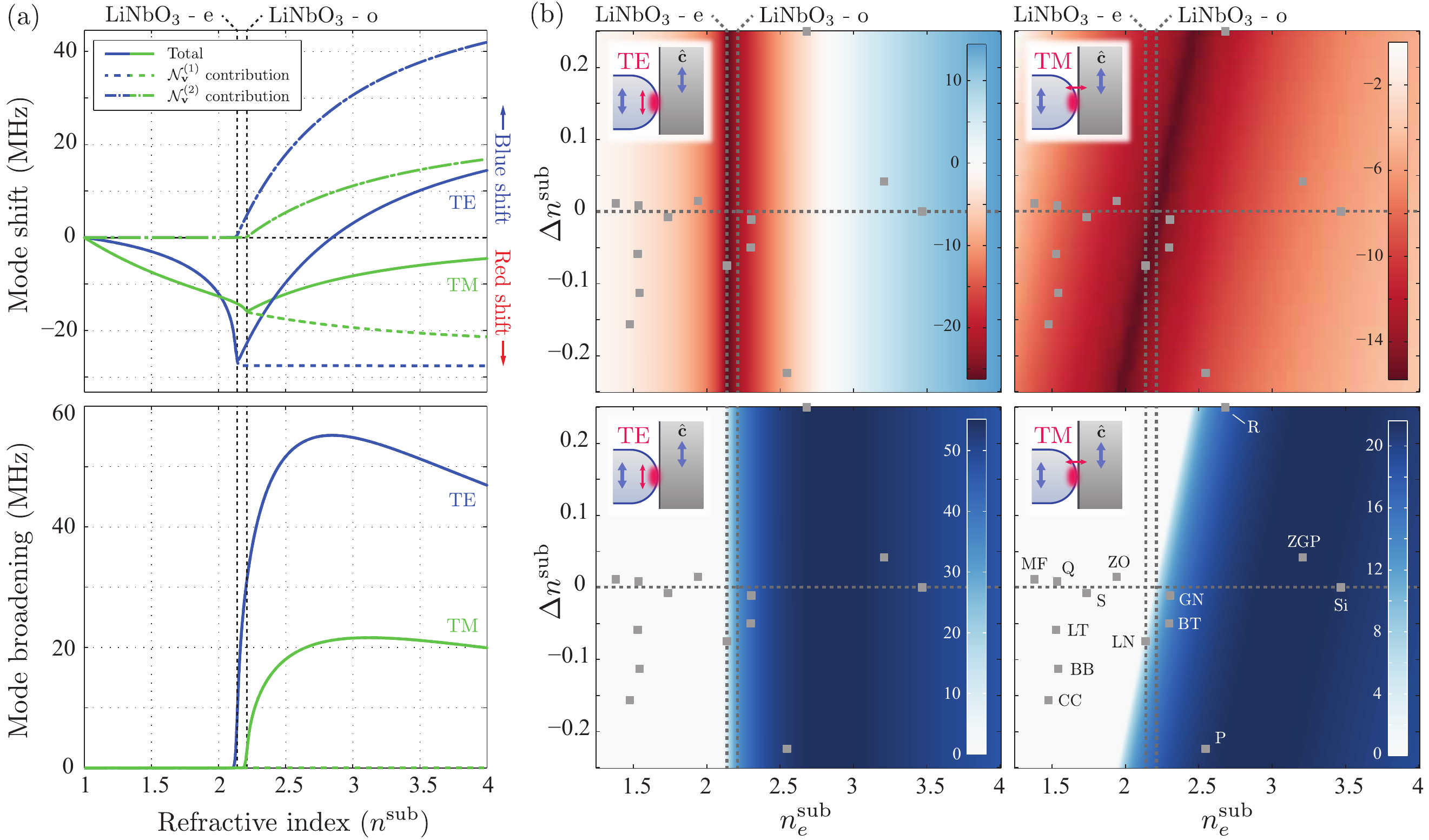}
		\caption{\textbf{Effects of substrate birefringence}: (a)
			Dependence of the resonance shift $\Delta \omega_{\mathbf{v}}/(2\pi)$ (top) and
			mode broadening $\Delta \gamma_{\mathbf{v}}/(2\pi)$ (bottom)
			in MHz of TE (blue)
			and TM (green) WGMs supported in a LiNbO$_3$ resonator
			induced by bringing an isotropic substrate of varying
			refractive index $n^{\txtpow{sub}}$ into contact. Total resonance
			modifications (solid lines) derive from both a near field material
			polarisation $\mathcal{N}_{\mathbf{v}}^{(1)}$ (dashed) and 
			a far field radiative $\mathcal{N}_{\mathbf{v}}^{(2)}$
			(dash-dotted)
			contribution. (b) Variation of the (top) mode shift (in MHz) and
			(bottom) broadening (in MHz) of TE (left) and TM (right) WGMs with substrate birefringence $\Delta
			n^{\txtpow{sub}} = n_e^{\txtpow{sub}} -
			n_o^{\txtpow{sub}}$. The optic axis of the substrate is
			assumed to be parallel to that of the resonator as depicted
			in the insets. Grey squares indicate material properties of some common
			(birefringent) substrates: magnesium
			fluoride, MgF$_2$ (labelled MF); crystal quartz, SiO$_2$ (Q); sapphire,
			Al$_2$O$_3$ (S); lithium
			tetraborate, Li$_2$B$_4$O$_7$ (LT); barium borate,
			BaB$_2$O$_4$ (BB); lithium niobate, LiNbO$_3$ (LN); calcite,
			CaCO$_3$ (C); zinc oxide, ZnO (ZO); gallium nitride, GaN
			(GN); barium titanate, BaTiO$_3$ (BT); proustite,
			Ag$_3$AsS$_3$ (P); zinc germanium phosphide ZnGeP$_2$ (ZGP); rutile, TiO$_2$ (R); and silicon, Si (Si). Dashed vertical lines depict the
			ordinary and extraordinary refractive indices of the
			resonator. Refractive index data were taken from \cite{Zelmon1997, Ghosh1999, Dodge1984, Eimerl1987, Sugawara1998,
				Jeppesen1958, Devore1951, Bond1965, Barker1973, Babicheva2013, Hulme1967, Das2003}. \label{fig:birefringence}}
	\end{center}
\end{figure*} 
Before considering the more general scenario of uniaxial substrates we
first analyse the 
resonance perturbations induced by an isotropic substrate. The solid curves in Figure~\ref{fig:birefringence}(a) show the calculated
resonance shift $\Delta \omega_{\mathbf{v}}/(2\pi)$ (top plot) and
mode broadening $\Delta \gamma_{\mathbf{v}}/(2\pi)$  (bottom) in MHz that result
when a dielectric substrate of varying refractive index,
$n^{\txtpow{sub}}$, is brought from infinity into contact with the
resonator ($h=0$). As described by \eqref{eq:final_shift} the total
complex frequency shift derives from two distinct contributions. Specifically, the
$\mathcal{N}_{\mathbf{v}}^{(1)}$ term in \eqref{eq:final_shift}
relates to work done in polarising the dielectric and to material
absorption therein (the latter is assumed to be zero in all calculations). The second
$\mathcal{N}_{\mathbf{v}}^{(2)}$ term, however, is required when
describing open cavities and relates to coupling of the WGM
into the far field \cite{Ruesink2015}, which can be further facilitated by introduction of
the substrate. The individual contributions from these terms are also
indicated in Figure~\ref{fig:birefringence}(a) by the dashed and
dot-dashed curves respectively. 

As the refractive index of the substrate increases from unity, it is
seen from Figure~\ref{fig:birefringence}(a) that the WGM resonance
frequency is red-shifted. For small refractive indices,
the transmitted field in the substrate is purely evanescent, such
that the red-shift originates solely from the work done generating a
material polarisation, which can equivalently be considered as an
increase in the effective refractive index, and hence optical path
length, of the WGM propagating in the resonator. Accordingly, no additional radiative losses are
introduced into the system and no mode broadening is seen. As
$n^{\txtpow{sub}}$ increases further so too does the
magnitude of the red shift (due to the larger resulting index contrast), until the refractive index of the
substrate is approximately equal to the effective refractive index of
the unperturbed WGMs i.e. $n^{\txtpow{sub}} \approx
n^{\txtpow{eff}}$ (albeit not exactly due to the non-zero width of the
WGM angular spectrum). Once this condition is met a maximum red shift
results. We note that by virtue of the
anisotropy of the lithium niobate resonator assumed in our
calculations, the effective refractive index for TE and TM WGMs
differs ($n^{\txtpow{eff}} \approx n_e$ and $n_o$ respectively) and
hence so too does the position of the maximum red shift. 

For yet larger values of $n^{\txtpow{sub}}$ the magnitude of
the red shift starts to decrease.  Physically, the qualitative change
in the frequency shift arises because when $n^{\txtpow{sub}} \gtrsim n^{\txtpow{eff}}$
the field transmitted into the substrate is no longer purely
evanescent, but can contain significant (or even only) propagating
components. Consequently, the total shift derives from competition
between the red shifts arising from the
$\mathcal{N}_{\mathbf{v}}^{(1)}$ term and blue shifts from the
radiative $\mathcal{N}_{\mathbf{v}}^{(2)}$ term. For (predominantly $s$
polarised) TE modes, the $\mathcal{N}_{\mathbf{v}}^{(1)}$ contribution is constant regardless of substrate refractive index since
the change in the material polarisability is precisely compensated for
by the change in transmission into and reflection from the substrate. The differing
dependence of the Fresnel coefficients on the substrate
refractive index for incident $p$ polarised
TM modes, however, means that the resulting red shift arising
from material polarisation weakly increases with
$n^{\txtpow{sub}}$. In contrast, the $\mathcal{N}_{\mathbf{v}}^{(2)}$
resonance shift, which arises from interference between the WGM and
the additional radiated field induced by shifting the position of the
substrate (or equivalently a back action from the radiation continuum
\cite{Ruesink2015}) is towards higher frequencies. Ultimately when the substrate refractive index is
large enough, the radiation interaction can dominate the resonance shift giving rise to an
anomalous net blue
shift of the WGM relative to the case when no substrate is
present. This transition from a red to blue shift occurs at
much lower substrate refractive indices for TE modes than for TM modes
due to the differing behaviour of $\mathcal{N}_{\mathbf{v}}^{(1)}$ for
$n^{\txtpow{sub}} \gtrsim n^{\txtpow{eff}}$. Practically, the
substrate refractive index required to observe blue shifts of TM modes
is unphysically large within the optical domain. Although we
shall not consider this case in any detail in this article, it is worth noting that WGM blue
shifts can also be observed when the refractive index of the medium
surrounding the resonator is greater than that of the substrate as follows
from the $\delta{\tensor\epsilon}$ dependence of \eqref{eq:mathcalN1},
or equivalently the change in sign of the material polarisability as
follows from the Clausius-Mossotti relation. 

\begin{figure*}[t!]
	\begin{center}
		\includegraphics[width=\textwidth]{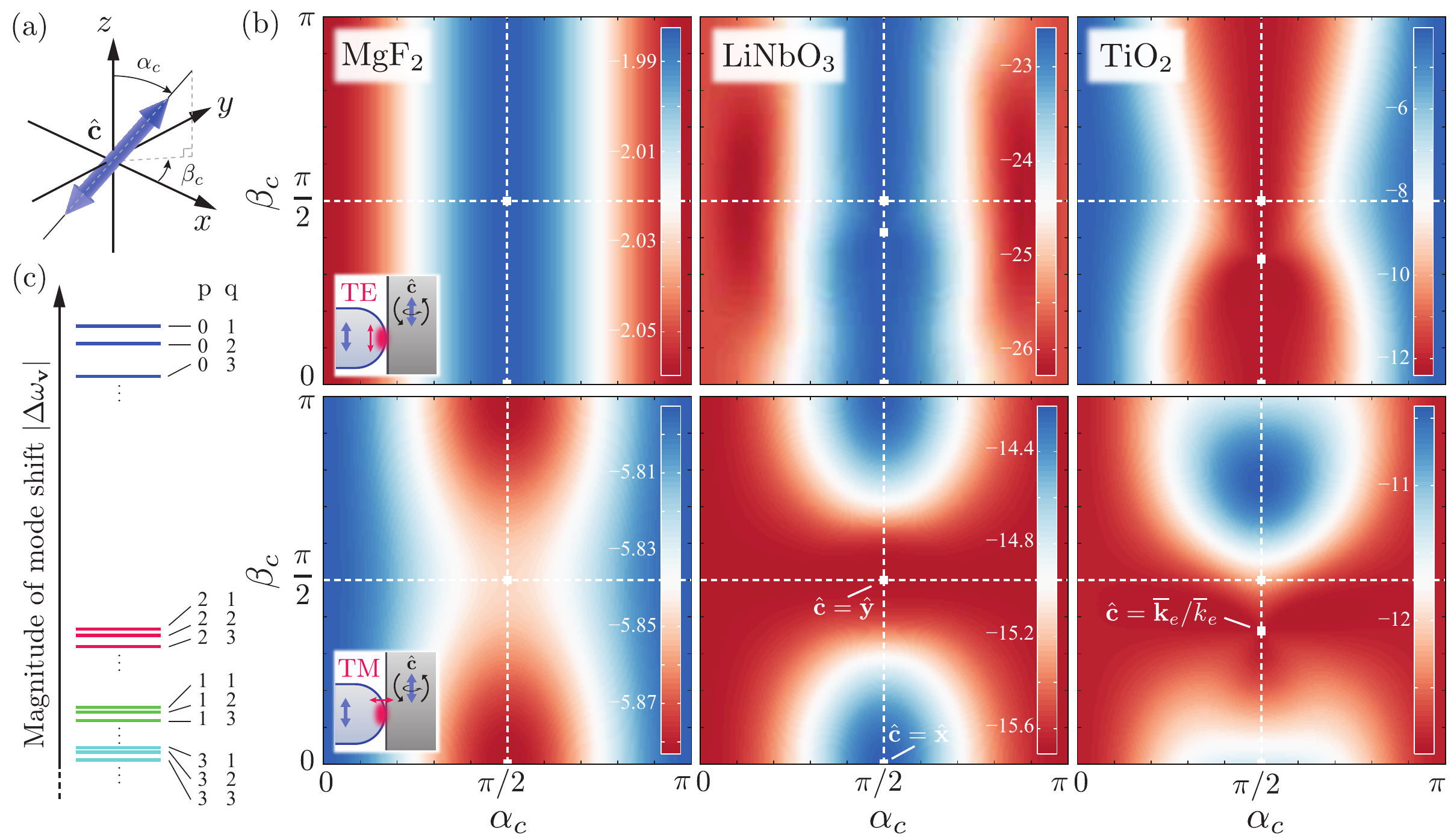}
		\caption{\textbf{Optic axis and modal dependence of resonance
				shifts}: (a) Coordinate system defining orientation of
			the optic axis $\hat{\mathbf{c}}$ of the substrate. (b)
			Resonance shifts $\Delta \omega_{\mathbf{v}}/(2\pi)$ (in MHz) induced in TE (top
			row) and TM (bottom) WGMs upon bringing a magnesium
			fluoride, MgF$_2$ (left column); lithium niobate, LiNbO$_3$
			(middle); or rutile, TiO$_2$ substrate into contact with a
			LiNbO$_3$ resonator, as a function of the orientation of the
			substrate optic axis $\hat{\mathbf{c}} = [\sin\alpha_c\cos\beta_c,
			\sin\alpha_c \sin\beta_c,\cos\alpha_c]$. Note that for the
			substrate refractive indices considered only red shifts are predicted. (c) Schematic of
			the typical hierarchy of resonances shifts for WGMs of fixed $(m,\nu)$ but
			differing polar and radial index $(p,q)$. A similar
			hierarchy is also seen for modes of (approximately) fixed
			wavelength, but differing $(m,p,q)$. \label{fig:shift}}
	\end{center}
\end{figure*} 
As noted above, when the refractive index of
the substrate is larger than the effective refractive index seen by
the WGM in the unperturbed resonator, light couples to radiative,
i.e. propagating waves in the substrate. Energy is carried by these
propagating waves to infinity such that they constitute a 
loss mechanism. Accordingly, once coupling of the WGM to these
propagating modes is possible, the line width of the resonance
increases as seen in Figure~\ref{fig:birefringence}(a). Briefly noting that the relative fraction of the mode
energy that is contained exterior to the resonator (\eqref{eq:Uext}) is
smaller for TM modes as compared to TE modes it would be expected that
the resulting mode broadening is also smaller for TM modes. This
expectation is indeed borne out in calculations as apparent from
Figure~\ref{fig:birefringence}(a). Furthermore, we observe that as we
increase the substrate refractive index, the coupling rate 
increases to a maximum value, before slowly decreasing at larger
refractive indices. This behaviour can be understood by first noting that
the Fresnel transmission coefficients decrease with increasing substrate
refractive index (i.e. there is stronger reflection at the
interface). Larger substrate refractive index, however, also implies that the $k_{x}$
component of the transmitted wavevector increases (i.e. the transmitted wave
propagates at a smaller angle to the surface normal), such that the
projection factor appearing when considering energy
conservation at the interface also increases.  These two opposing effects give rise
to the maximum seen in Figure~\ref{fig:birefringence}(a).

Extension of the calculations to the case of uniaxial substrates
shows similar trends as to the isotropic case, as shown in
Figure~\ref{fig:birefringence}(b). Specifically, the resonance shift $\Delta\omega_{\mathbf{v}}/(2\pi)$
(top row) and mode broadening $\Delta\gamma_{\mathbf{v}}/(2\pi)$ (bottom row) in MHz are shown as a
function of the extraordinary refractive index of the substrate
$n_e^{\txtpow{sub}}$ and the substrate birefringence $\Delta
n^{\txtpow{sub}} = n_e^{\txtpow{sub}} - n_o^{\txtpow{sub}}$. The optic
axis of the substrate is assumed to be parallel to that of the
resonator, i.e. $\hat{\mathbf{c}} = \hat{\mathbf{z}}$. Although a weak dependence of the TE shift and broadening on the
substrate birefringence results from the cross-polarisation mixing
described in Eqs.~\myeqref{eq:Eref}--\myeqref{eq:Eref2}, this effect 
is found to be several orders of magnitude smaller than the co-polarised terms, such that only
variation of the resonance properties with the substrate extraordinary
index, and not $\Delta n^{\txtpow{sub}}$, is seen in 
Figure~\ref{fig:birefringence}(b), mirroring
that found for the isotropic case. When $\hat{\mathbf{c}} =
\hat{\mathbf{z}}$, TM modes also effectively see an isotropic
substrate with refractive index $n_o^{\txtpow{sub}}$. As such the
resulting functional dependence of the shifts and broadening exhibits
a linear displacement with substrate birefringence for fixed $n_e^{\txtpow{sub}}$. This effect
could, for example, hence be exploited for differential tuning of TE and TM
WGMs if the substrate has a large linear electro-optic coefficient and
the resonator itself does not show strong electro-optic effects. 

Anisotropy
of the substrate also plays an important role in
dictating the WGM shifts when the optic axis is 
varied, as shown in Figure~\ref{fig:shift}. Letting $\hat{\mathbf{c}}
= [\sin\alpha_c \cos\beta_c, \sin\alpha_c\sin\beta_c,\cos\alpha_c]$,
where $\alpha_c$ and $\beta_c$ are the polar and azimuthal angles of
the optic axis (see Figure~\ref{fig:shift}(a)), we have calculated the
resulting resonance shift when a magnesium fluoride (MgF$_2$: $n_e = 1.382$, $n_o = 1.371$), lithium
niobate (LiNbO$_3$: $n_e = 2.138$, $n_o =
2.213$) or rutile (TiO$_2$: $n_e = 2.683$, $n_o = 2.432$) prism is brought into contact
with the LiNbO$_3$ resonator from infinity, as a function of
$(\alpha_c,\beta_c)$. These three specific materials were selected
since both the ordinary and extraordinary refractive index of 
MgF$_2$ (TiO$_2$) are smaller (larger) than that of the LiNbO$_3$
resonator, whereas choice of a LiNbO$_3$ substrate also allows the
intermediate regime to be analysed. Numerical results are shown in Figure~\ref{fig:shift}(b).

Considering first the MgF$_2$ substrate, we note that since all
waves transmitted into the substrate are evanescent the resonance
shift originates only from the work done in polarising the medium as
before. Accordingly, since MgF$_2$ is positively birefringent 
($n_e^{\txtpow{sub}}>n_o^{\txtpow{sub}}$), if the optic axis is rotated from the
$\hat{\mathbf{z}}$ axis towards the $x$-$y$ plane ($\alpha_c$ is increased) the magnitude of the shift of the TE mode decreases because the
effective refractive index seen by the WGM also decreases. Conversely,
since the TM WGM is predominantly polarised in the $x$-$y$ plane a larger
magnitude resonance shift results under the same rotation of
$\hat{\mathbf{c}}$. Rotation of the optic axis around the
$\hat{\mathbf{z}}$ axis through an increase of $\beta_c$, also gives rise
to a modulation in the resonance shift of the TM mode through the same
mechanism. The largest magnitude shift therefore results when the
optic axis is perpendicular to the field component with the greatest
amplitude in the near field. For TM modes this corresponds to $\beta_c
= 0$ i.e. the radial direction at the contact point between the
resonator. No variation in the TE resonance frequency with $\beta_c$
is however seen, since the mode is polarised along the axis of rotation.
For a negatively birefringent substrate ($n_e^{\txtpow{sub}} <
n_o^{\txtpow{sub}}$), such as barium borate, these trends would be
reversed, e.g. the TE mode would experience
smaller red shifts as $\alpha_c$ was increased to $\pi/2$ as opposed
to larger shifts. We also note that the modulation of the resonance
shift due to the variation of the material polarisability increases
with the birefringence of the substrate. A rough order of magnitude
estimate of the relative modulation can be obtained by considering the
relative change in the Clausius–Mossotti polarisability when evaluated
using the ordinary and extraordinary refractive indices. Typically,
for the materials considered in this work variation of only a few
percent (or equivalently $\lesssim 1$~MHz) were seen as also evidenced by the data in Figure~\ref{fig:shift}.

Although the dependence of the resonance shifts for WGMs
resulting from the presence of a TiO$_2$ substrate are qualitatively
similar to those for a MgF$_2$ substrate, the physical origin is quite
different. This is apparent since both TiO$_2$ and MgF$_2$ are
positively birefringent materials, yet the observed trends are in
opposition to each other. Since
both the ordinary and extraordinary refractive indices of TiO$_2$ are
larger than that of the resonator, the field in the substrate is
composed of propagating waves. As was discussed above, as the optic
axis of the substrate is varied, so the effective refractive index of
the substrate seen by the WGM also varies, however, with reference to
Figure~\ref{fig:birefringence}(a), when in the propagating transmitted
wave regime ($n_{e,o}^{\txtpow{sub}} \gtrsim n^{\txtpow{eff}}$) the
relative change in the resonance frequency is dominated by the
radiative shift term ($\mathcal{N}^{(2)}_{\mathbf{v}}$). Mathematically speaking, since
$|\partial\, \mbox{Re}[i\mathcal{N}^{(2)}_{\mathbf{v}}] / \partial
\hat{\mathbf{c}} | > |\partial \, \mbox{Re}[\overline{\omega}\mathcal{N}^{(1)}_{\mathbf{v}} ]/ \partial
\hat{\mathbf{c}} |$ when $n_{e,o}^{\txtpow{sub}} \gtrsim
n^{\txtpow{eff}}$, the optic axis dependence of the resonance shift is
dominated by the radiative component of the Bethe-Schwinger
equation. Importantly, negative
frequency shifts can still be observed (as in the TiO$_2$ data in
Figure~\ref{fig:shift}(b)) if
$|\mbox{Re}[\overline{\omega}\mathcal{N}^{(1)}_{\mathbf{v}}]| >
|\mbox{Re}[i\mathcal{N}^{(2)}_{\mathbf{v}}]|$. The modulation of the resonance
frequency from the radiative interaction corresponds to several tens
of percent, or equivalently to $\lesssim 10$~MHz, depending on the
material. Considering TiO$_2$ for definiteness, we note that as
the optic axis is rotated  from the
$\hat{\mathbf{z}}$ axis towards the $x$-$y$ plane ($\alpha_c$ is
increased) such that the effective refractive index seen by a TE mode
decreases, coupling of the WGM to the far field is reduced due to
lower transmission. Consequently the corresponding blue shift from the
radiative back action is also reduced such that in total a larger red
shift is observed (see Figure~\ref{fig:shift}(b)). The converse
again holds for TM modes. Interestingly, in the frequency shifts
calculated for rutile a weakly singular feature can also be discerned,
which occurs when the central wavevector of the output extraordinary
beam is parallel to the optic axis of the substrate. This special case will be discussed
further below. 

When a LiNbO$_3$ substrate is used to tune the resonance frequency a mixed
behaviour is seen. Due to the choice of optic axis of the resonator, the
refractive index experienced by a TE WGM in the substrate is always
larger than the effective refractive index in the
resonator. Accordingly a propagating field is transmitted into the
substrate and the dependence on the optic axis is dictated by the
variation in $\mathcal{N}_{\mathbf{v}}^{(2)}$ as was the case for
TiO$_2$ (we note the opposing trends between LiNbO$_3$ and TiO$_2$
is a result of the opposite signature of the birefringence). TM modes, 
however, couple predominantly to evanescent waves in the substrate and
hence the dependence on the optic axis is governed by
$\mathcal{N}_{\mathbf{v}}^{(1)}$ as was found with MgF$_2$.

In the calculations presented thus far we have only
considered the resonance shifts induced in WGMs for which
$p = 0$ and $q = 1$. Naturally, a quantitative difference in the
induced shift is seen for modes of differing polar and radial
order (but fixed $m$ and $\nu$). Under typical conditions the hierarchy
of shifts is that shown schematically
in Figure~\ref{fig:shift}(c). Specifically, as the radial order
increases, so the magnitude of the shift falls slightly. Modes of
differing polar order, however, form two separate
ladders corresponding to modes of the same
parity (i.e. odd or even symmetry). Within each individual ladder, smaller magnitudes shifts are
seen for higher values of $p$ because for such higher order modes a
greater proportion of the mode lies outside the Gaussian coupling
window and hence does not interact significantly with the substrate. Deviations from the mode
ordering shown in Figure~\ref{fig:shift}(c) do, however, occur when
the shifts are close to zero. It should
also be noted that since modes of fixed $m$ have been considered here,
modes of differing ($p,q$) have different resonant
frequencies. Nevertheless, if WGMs of approximately fixed wavelength
are chosen (and $m$ varied accordingly) a similar ordering is also found.

\subsection{Selective coupling}
\begin{figure*}[!p]
	\begin{center}
		\includegraphics[width=\textwidth]{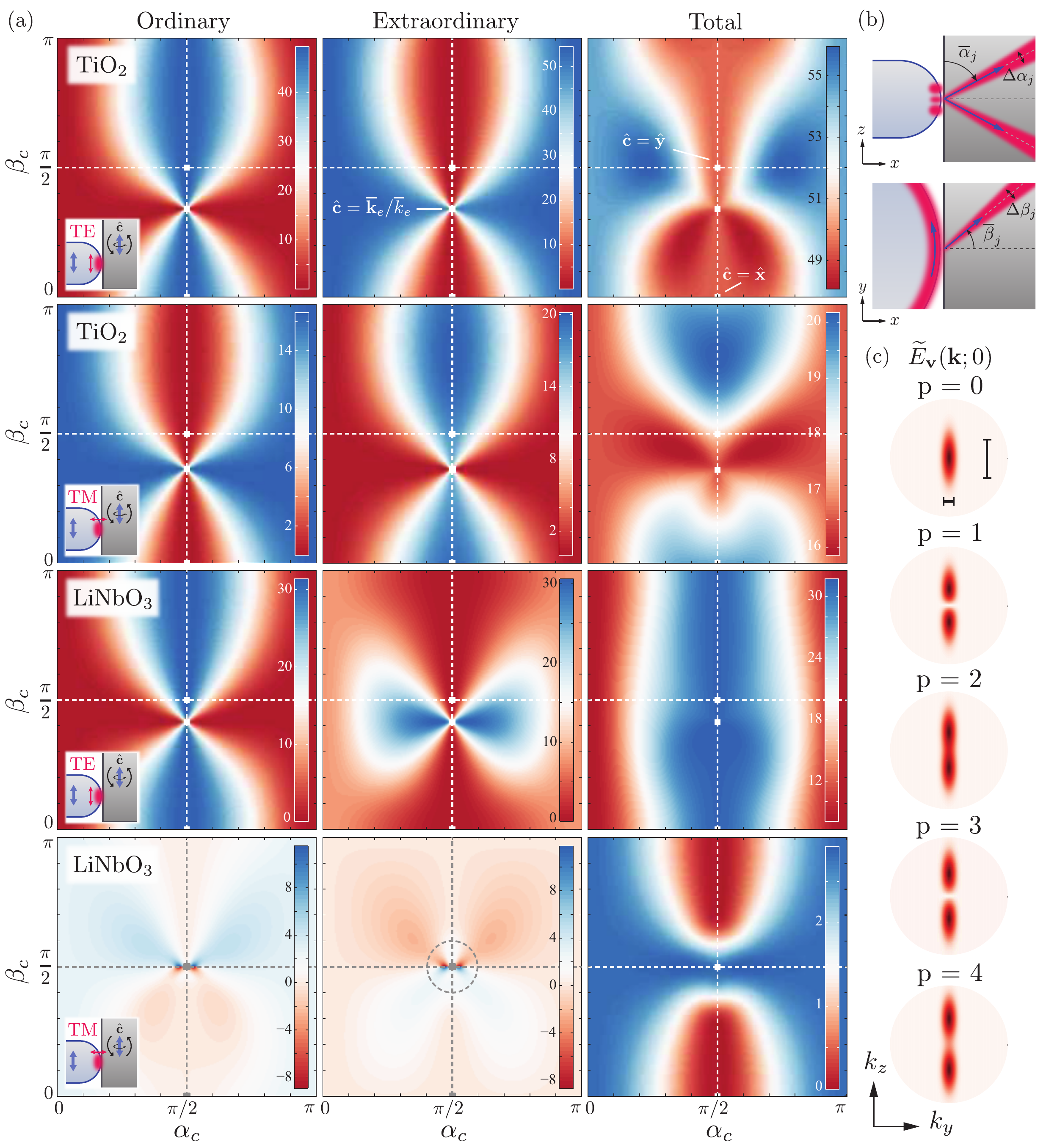}
		\caption{\textbf{Mode broadening and out-coupled beam parameters}: (a) Mode broadening $\Delta\gamma_{\mathbf{v}}/(2\pi)$ (in MHz) of TE (first
			and third rows) and TM (second and fourth rows) WGMs upon
			bringing a TiO$_2$ (top two rows) or LiNbO$_3$ (bottom two
			rows) substrate  into contact with a
			LiNbO$_3$ resonator, as a function of the orientation of the
			optic axis $\hat{\mathbf{c}}$ of the substrate. The total
			mode broadening (right column) results from losses carried
			by an out coupled ordinary (left) and extraordinary (middle)
			beam. (b) Schematic defining the angular direction and
			angular width of propagating beams coupled into the
			substrate. (c) Scalar profiles
			$\widetilde{E}_{\mathbf{v}} (\mathbf{k},0)$ of 
			out-coupled beams of different polar order $p$. Scale bars
			correspond to the FWHM of the fundamental $p=0$ mode in the
			$k_y$ and $k_z$ directions i.e. $2\sqrt{2\ln 2}/\Delta y$ and $2\sqrt{2\ln
				2}/\Delta z$ respectively.  \label{fig:broad}}
	\end{center}
\end{figure*}

In addition to the resonance shifts discussed in the previous section,
introduction of a substrate can give rise to coupling of the WGM into
the far field and hence to  mode
broadening. Intuitively, this phenomenon can be understood since 
coupling to propagating
waves in the substrate introduces additional loss channels for the
WGM, yielding a shorter resonance lifetime. Since propagating waves are only excited when
the effective refractive index of the substrate is larger than that
seen by the WGM in the resonator, we now consider only the effect of
either a
LiNbO$_3$ or TiO$_2$ substrate. Figure~\ref{fig:broad}(a) shows the
mode broadening determined using \eqref{eq:final_shift} as a
function of the orientation of the optic axis of the
substrate. Anisotropy of the substrate implies that a given WGM can
couple to both ordinary and extraordinary polarised waves in the substrate, such
that the mode broadening results from the net effect of both
channels. Accordingly, we also show the individual contributions from
the ordinary and extraordinary beams in the leftmost columns of Figure~\ref{fig:broad}(a).

Initially considering a TiO$_2$ substrate, we again note that as
$\alpha_c$ is increased to $\pi/2$ the refractive index seen by TE ($s$ polarised)
WGMs decreases. Lower transmission into the far field hence results such
that the mode broadening is reduced. Moreover, as the optic axis is
rotated the ordinary and extraordinary polarisation vectors also
rotate, such that the ordinary and extraordinary waves play complementary
roles. For instance, when $\hat{\mathbf{c}} = \hat{\mathbf{z}}$ the TE
mode couples to only extraordinary waves in the substrate,
however, if $\alpha_c=\pi/2$ a strong ordinary component can also be
excited. Proportionally, the relative strength with which TE modes
couple to ordinary waves varies with $\beta_c$. A weakly singular point is seen in the plots of
Figure~\ref{fig:broad}(a) corresponding to the case where
$\hat{\mathbf{c}} = {\overline{\mathbf{k}}_e}/|\overline{\mathbf{k}}_e|$. When this condition is meet
there is a minimum (maximum) in the refractive index experienced by the central
angular component of the extraordinary
wave when propagating in a positively (negatively) birefringent
material. Furthermore, the polarisation vector $\hat{\mathbf{e}}$ is
undefined for this angular component, such that a plane wave component
propagating with wavevector parallel to $\hat{\mathbf{c}}$ does so as
a pure ordinarily polarised wave. Nevertheless, for an out-coupled WGM
the extraordinary beam still carries a finite (albeit relatively
small) amount of power to the
far field due to the finite width of the incident angular spectrum. 
For an isotropic substrate, the ordinary and extraordinary wave
vectors are equal and the complementarity of the corresponding beams
is exact in the sense that a decrease in the energy carried by the
ordinary beam resulting from a change in $\hat{\mathbf{c}}$, is
compensated by an increase of equal magnitude in that of the extraordinary wave. Quantitative differences arise for
anisotropic substrates however (as seen in Figure~\ref{fig:broad}(a)),
due to the differing dependence of the generalised Fresnel transmission
coefficients and since $\mathbf{k}_o$ and $\mathbf{k}_e$ are not
parallel. Similar, albeit reversed arguments apply for a TM WGM coupled to a TiO$_2$ substrate. When broadening of WGMs of differing
polar and radial orders is considered, a similar hierarchy to that
for resonance shifts is again found (see Figure~\ref{fig:shift}(c)).

When a LiNbO$_3$ substrate is brought into close proximity to a WGM
resonator, markedly different behaviour can be seen. As noted
in the previous section, a TE WGM always couples to propagating waves in the
substrate yielding similar behaviour to that seen for a TiO$_2$
substrate (allowing for differences arising from the sign of the
birefringence of each material). In contrast a TM mode mostly couples to evanescent waves except
for within a restricted range of $\hat{\mathbf{c}}$ corresponding to
that bounded by the grey dashed circle in
Figure~\ref{fig:broad}(a). Only
relative weak broadening of the TM mode hence results. Dependence of
the partial mode broadening, originating from the ordinary and
extraordinary waves individually, on $\hat{\mathbf{c}}$ is in contrast
to the propagating wave case dominated by the
non-radiative component ($\mathcal{N}^{(1)}_{\mathbf{v}}$) of the
Bethe-Schwinger equation since the radiative losses are so
low. Importantly, net mode broadening of TM WGMs perturbed by a LiNbO$_3$ substrate
is nevertheless governed by the radiative loss
($\mathcal{N}^{(2)}_{\mathbf{v}}$) term as per physical intuition,
since the near field  ($\mathcal{N}^{(1)}_{\mathbf{v}}$) contributions
from the ordinary and extraordinary waves are equal and opposite in
magnitude and therefore precisely sum to zero. 
We briefly note, that the sudden transition from mode broadening to
mode narrowing (and vice-versa) seen at $\beta_c = \pi/2$ in the
individual loss channels is an artifact of the definition of the
(extra-)ordinary polarisation and wave vectors and is not of physical
significance, especially considering there is no net effect from both channels.
Given that a LiNbO$_3$ substrate produces significant broadening of TE, but not TM, modes we note that this scenario corresponds
to polarisation selective coupling of WGMs as has recently been
experimentally demonstrated \cite{Sedlmeir2016}. Maximum extinction
ratios, that is to say the relative coupling rates, are achieved when $\hat{\mathbf{c}}$ lies in the
$x$-$y$ plane ($\alpha_c = \pi/2$), specifically along the
$\hat{\mathbf{x}}$ axis ($\beta_c = 0$). In this
case extinction ratios $\sim 1.6\times 10^4$ (or equivalently $\sim 42$~dB) are predicted corresponding to a
strong degree of selectivity.

\subsection{Geometric properties of out-coupled modes}
Finally, in this section we briefly consider the geometric properties of out coupled
propagating ordinary and extraordinary beams in a uniaxial substrate
when it is brought close to a WGM resonator. Specifically we seek
expressions for the output angle of the beams and the angular
widths as defined in Figure~\ref{fig:broad}(b), which are important
parameters in mode identification \cite{Schunk2014}. Similar considerations
for the emission patterns of WGMs supported in spheroidal resonators
coupled using isotropic prisms were given by Gorodetsky and Ilchenko
\cite{Gorodetsky1994}, however, here we provide the corresponding
expressions for a uniaxial prism. Most importantly, it should be noted
that use of a uniaxial prism means that the angular components of the
WGM each undergo double refraction, such that two beams are output in
different directions and with differing polarisation state. Depending
on the magnitude of the substrate birefringence and optic axis
orientation, these two beams can, however, partially or completely overlap.

We begin by first writing the wavevector of a
general angular component in the form $\mathbf{k} =
k[\sin\alpha\cos\beta, \sin\alpha \sin\beta,\cos\alpha]$ (see Figure~\ref{fig:geometry}).  Noting that the
transverse wave components of each angular component are conserved
across the substrate interface we can deduce the output angle in the
$x$-$y$ plane ($\overline{\beta}_j$) by considering $\overline{k}_y = k_{yr} = m/R = k_j
\sin\overline{\alpha}_j\sin\overline{\beta}_j$ for $j \in \{o,e\}$. It immediately follows that
\begin{align}
\sin\overline{\beta}_j = \frac{m}{k_j R \sin\overline{\alpha}_j}.\label{eq:beamparam1}
\end{align}
The output angles in the $z$ direction require a slightly more
detailed analysis. Within the Gaussian window the scalar mode profile
incident on the interface is an oscillatory function such that it
follows that, except for
the trivial case of $p=0$ for which $\overline{\alpha}_j = \pi/2$, two
distinct peaks located symmetrically about $k_z=0$ are expected in
the far field (as shown in Figures~\ref{fig:broad}(b) and
(c)). As an approximation we first consider the $k_z$ dependence of 
the incident angular
spectrum neglecting the effect of the Gaussian coupling
window which can be easily shown to be given by the Hermite
functions. To first order, the output beams correspond to the outermost
peaks of the Hermite functions, which can be well approximated as
\begin{align}
\exp[-u^2/2] H_p(u) \sim 2^{(2p+1)/4} \sqrt{\pi (p-1)!}
\, \mbox{Ai}\left[t\right] ,
\end{align}
where $t = 2^{1/2} p^{1/6} [(2p+1)^{1/2} - u]$ and $u = P \Theta_m
k_z$. Accordingly, the peaks are found at angles given by
\begin{align}
\cos\overline{\alpha}_j \approx \frac{1}{k_j  P \Theta_m} \left[\sqrt{2p+1}
- \frac{\zeta_1'}{2^{1/2}p^{1/6}}\right],
\end{align}
where $\zeta_1' = 1.0188$ is the first zero of the derivative Airy
function $\mbox{Ai}'[-\zeta] = 0$.
In the far field the Gaussian coupling window is accounted for by a
convolution of the Hermite functions with a Gaussian (as discussed
in Section~\ref{sec:pert_modes}). The asymmetry of the outermost peak of the Hermite functions,
however, gives rise to an additional angular displacement in the final
output beam, whereby we numerically find that
\begin{align}
\cos\overline{\alpha}_j &\approx \frac{1}{k_j  P \Theta_m} \left[\sqrt{2p+1}
- \frac{\zeta_1'}{2^{1/2}p^{1/6}}\right] \nonumber\\
&\quad\quad\quad+
\frac{1}{4}\left[\frac{\kappa}{P} + \frac{1}{P^2\Theta_m^2}\right]^{1/2}.\label{eq:beamparam2}
\end{align}
Determination of the angular widths of the output beams follows by
making the approximation $\Delta k_y / \Delta \beta_j \approx
| \partial k_y/\partial \beta|_{\alpha=\overline{\alpha}_j,\beta =
	\overline{\beta}_j }$ (and similarly for $\Delta k_z/\Delta
\alpha_j$). Within this approximation we find 
\begin{align}
\Delta \alpha^2_j &= \frac{k}{k_j^2} \left[\delta_{p,0}
\frac{n}{\sqrt{PR}} + \frac{\sqrt{n^2 -1}}{P \sin^2 \overline{\alpha}_j}\right],\label{eq:beamparam3}\\
\Delta\beta_j^2 &= \frac{k\sqrt{n^2-1}}{k_j^2 R \sin^2
	\overline{\alpha}_j\cos^2\overline{\beta}_j} .\label{eq:beamparam4}
\end{align}
Although similar in form to the expressions of Gorodetsky and Ilchenko,
Eqs.~\myeqref{eq:beamparam1}, \myeqref{eq:beamparam3} and
\myeqref{eq:beamparam4} possess slight differences due to differing
definitions of the angles and the shape of resonators
considered.  \eqref{eq:beamparam2} however differs from that of \cite{Gorodetsky1994} due to
the method of derivation, although we find \eqref{eq:beamparam2} gives
good numerical agreement with the positions of the peaks. We emphasis, however, that the output beam
profiles shown in Figure~\ref{fig:broad}(c) are further modulated by
the Fresnel transmission coefficients. Due to the narrow width of the
angular spectrum the associated variation of the transmission
coefficients is small, however, if
coupling to either ordinary or extraordinary propagating waves in the
substrate is close to cut-off, then strong variation can occur
resulting in so-called Fresnel filtering \cite{Rex2002,Tureci2002} and a truncated output beam profile.

\section{Conclusions}\label{sec:conclusions}

In this article we have theoretically studied the properties of WGMs in a $z$-cut uniaxial axisymmetric resonator placed in close proximity
to a planar dielectric substrate, such as a prism. To do so, we first
extended the approximate unperturbed mode profiles provided by Breunig et
al. \cite{Breunig2013} to account for the open nature of the resonator
and the vectorial nature of the underlying electromagnetic fields. Upon establishing an
analytic angular spectrum representation of WGMs
(\eqref{eq:angspec}) we proceeded to study prism induced coupling to other
WGMs and rigorously proved that coupling between modes of differing azimuthal
order and parity was forbidden. Within a resonant mode approximation
we subsequently derived expressions for the WGM profiles in the
presence of a substrate, including reflection from and transmission
through the interface
(Eqs.~\myeqref{eq:Er_approx}, \myeqref{eq:Et_NF_TETM_app} and
\myeqref{eq:Et_NF_TETM_app2}). We have furthermore presented an
analytic formalism capable of quantitatively predicting the resonance
shifts and mode broadening that result from bringing a planar
dielectric substrate from infinity into the evanescent field of a
WGM (\eqref{eq:final_shift}). Our analysis was based on a 
generalisation of the Bethe-Schwinger perturbation equation which
accounts for far-field radiative contributions and is hence suitable
for application to open cavities \cite{Ruesink2015}. Importantly,
through use of generalised Fresnel reflection and transmission
coefficients our theory can account for uniaxial substrates with an arbitrarily oriented optic
axis and, moreover, is applicable to disc, toroidal and ellipsoidal
shaped resonators, therefore extending alternative coupling
theories that can be found in the literature \cite{Gorodetsky1999} to
previously unconsidered geometries. 

To support our theoretical developments, extensive numerical examples
were also presented. Competing red and blue shifts were
found to dictate the total change in resonance frequency. The former red shift is well known
and derives from the work done in polarising the dielectric
substrate, however, the second less familiar effect originates from 
interference between the unperturbed and perturbed radiative component
of the WGM in the far field, or a so-called radiative back action. Relative dominance of each effect was found to be
influenced by the both the mode polarisation and the refractive index of
the perturbing substrate relative to that of the resonator, since the
latter determines whether WGMs couple to 
evanescent or propagating waves in the substrate. For both TE and TM
WGMs, maximum red shifts occur when  the refractive index of the
perturbing substrate is approximately equal to that of the resonator,
however, for larger substrate indices, the radiative contribution
begins to dominate, leading not only to anomalous
blue shifts, but also to significant mode
broadening. An optimal regime, with respect to the substrate
refractive index, in which maximum coupling can be
achieved was also observed. Coupling of WGMs to the far field
by means of a uniaxial prism, however, can manifest itself through two
distinct channels, namely those of ordinary and extraordinary propagating waves. Orientation of
the optic axis plays a central role in governing the relative losses
between these channels. Exceptionally, when the resonator and substrate are
fabricated from the same (birefringent) material this dependence
renders polarisation selective coupling with a high extinction ratio
possible. Optimal selective coupling is found when the optic axis of
the resonator and substrate are chosen to lie parallel to the
resonator's symmetry axis ($z$-cut) and perpendicular to the substrate
interface ($x$-cut) respectively. 
Finally, we have discussed double refraction of a WGM into ordinary and
extraordinary waves propagating in a nearby anisotropic substrate and moreover
provided analytic formulae (Eqs.~\myeqref{eq:beamparam1} and \myeqref{eq:beamparam2}-\myeqref{eq:beamparam4}) for their geometrical properties under general conditions.

The authors would like to acknowledge A. Aiello, Ch. Marquadt, U. Vogl, G. Schunk
and D. Strekalov for helpful discussions.

\appendix

\section{Reflection and transmission at a birefringent interface
}\label{app:Fresnel}
In this appendix we derive the generalised Fresnel reflection and transmission 
coefficients of a plane wave in an isotropic medium incident upon a
planar uniaxial interface whose surface normal is directed along $\hat{\mathbf{x}}$. The incident wave is assumed to have a complex wavevector,
$\mathbf{k} = [k_x,k_y,k_z]$ (where
$(x,y,z)$ is a fixed coordinate system as used in the main text), and can be decomposed into $\hat{\mathbf{s}} =(\mathbf{k}\times \hat{\mathbf{x}})/|\mathbf{k}\times \hat{\mathbf{x}}|$ and
$\hat{\mathbf{p}} = (\mathbf{k}\times \hat{\mathbf{s}})/|\mathbf{k}\times \hat{\mathbf{s}}|$ polarised components which are 
perpendicular and parallel to the plane of incidence respectively. We further define the
unit vectors $\hat{\mathbf{h}} = \hat{\mathbf{s}}$, which is parallel to
the interface and perpendicular to the plane of incidence, and $\hat{\mathbf{g}} = \hat{\mathbf{x}}\times \hat{\mathbf{h}}$ which lies in the plane of incidence
and is parallel to the interface. The incident wave gives rise to a reflected
beam with wavevector $\mathbf{k}_r  = [-k_x,k_y,k_z]$ which can be decomposed into
$\hat{\mathbf{s}}_r = \hat{\mathbf{s}}$ and
$\hat{\mathbf{p}}_r = [\mathbf{k}_r\times \hat{\mathbf{s}}] /|\mathbf{k}_r\times \hat{\mathbf{s}}|$ polarised
components. Within a uniaxial medium, with arbitrary optic axis
$\hat{\mathbf{c}} = [c_x,c_y,c_z]$, both ordinary and extraordinary
wave components are generated with the associated unit polarisation
vectors (corresponding to the electric field vector)
\begin{align}
\hat{\mathbf{o}} &= (\mathbf{k}_o\times \hat{\mathbf{c}} )\, \big/\,
\left|\mathbf{k}_o\times \hat{\mathbf{c}} \right| \\
\hat{\mathbf{e}} &=  \tensor{\epsilon}{}^{-1} (\mathbf{k}_e \times [
\mathbf{k}_e \times \hat{\mathbf{c}} ]) \Big/
\left| \tensor{\epsilon}{}^{-1} \left(\mathbf{k}_e \times [\mathbf{k}_e \times \hat{\mathbf{c}} ]\right)\right| 
\end{align}
where 
\begin{align}
\tensor{\epsilon} = \!\left[\!\!
\begin{array}{ccc}
c_x^2 (n_e^2-n_o^2)+n_o^2 & c_x c_y (n_e^2-n_o^2) & c_x c_z (n_e^2-n_o^2) \\
c_x c_y (n_e^2-n_o^2) & c_y^2 (n_e^2-n_o^2)+n_o^2 & c_y c_z (n_e^2-n_o^2) \\
c_x c_z (n_e^2-n_o^2) & c_y c_z (n_e^2-n_o^2) & c_z^2 (n_e^2-n_o^2)+n_o^2 
\end{array}\!\!
\right]
\end{align}
is the permittivity tensor and $n_o$ ($n_e$) are the (extra)ordinary
refractive indices of the dielectric substrate (note that in the main
text these are denoted by $n_o^{\txtpow{sub}}$ and
$n_e^{\txtpow{sub}}$ respectively). The corresponding wavevectors are
$\mathbf{k}_o = [k_{o,x},k_y,k_z]$ and $\mathbf{k}_e =
[k_{e,x},k_y,k_z]$ where we note that the $k_y$ and $k_z$
components are conserved quantities across the interface (as per
Snell's law). Letting $\hat{\mathbf{c}} = [\cos\chi_c,\sin\chi_c\cos\psi_c,
\sin\chi_c\sin\psi_c]$, the $x$ components of the wavevectors are given by \cite{Gu1993a}
\begin{align}
k_{o,x} &= \sqrt{n_o^2 k^2 - k_y^2 - k_z^2}, \label{eq:kox}\\
k_{e,x}&= \left[v+\sqrt{v^2 - 4 u w}\right]/(2u)  \label{eq:kex}
\end{align}
and
\begin{align}
u &= \frac{\sin^2 \chi_c}{n_e} + \frac{\cos^2 \chi_c }{n_o}, \label{eq:udef}\\
v&= [k_y \cos \psi_c + k_z \sin\psi_c] \sin 2 \chi_c
\left(\frac{1}{n_e^2} + \frac{1}{n_o^2}\right),\label{eq:vdef}\\
w&= [k_y \cos \psi_c + k_z \sin\psi_c] ^2  \left(\frac{\cos^2 \chi_c}{n_e^2} +
\frac{\sin^2 \chi_c}{n_o^2}\right) \nonumber \\
&\quad\quad\quad\quad + \frac{[-k_y\sin\psi_c + k_z
	\cos\psi_c]^2}{n_e^2} - k^2.\label{eq:wdef}
\end{align}
The incident, reflected and
transmitted electric fields can then be written (omitting the $\exp(-i\omega t)$ time
dependence) as
\begin{align}
\mathbf{E}_i &= \left[A_s \hat{\mathbf{s}} +
A_p \hat{\mathbf{p}} \right] \exp[i  \mathbf{k}\cdot\mathbf{r}] \\
\mathbf{E}_r &= \left[B_s \hat{\mathbf{s}} +
B_p \hat{\mathbf{p}}_r \right]  \exp[ i \mathbf{k}_r\cdot\mathbf{r}]\\
\mathbf{E}_t &= C_o \hat{\mathbf{o}} \exp[i \mathbf{k}_o\cdot\mathbf{r}] +
C_e \hat{\mathbf{e}} \exp[ i\mathbf{k}_e\cdot\mathbf{r}],
\end{align}
where $A_{s,p}$, $B_{s,p}$, $C_{o,e}$ are amplitude coefficients and
$\omega$ is the optical frequency. From Maxwell's equation $i\omega
\mu \mathbf{H} = \nabla \times \mathbf{E}$, we can also express the
associated magnetic fields in the form:
\begin{align}
\omega \mu \mathbf{H}_i &= \left[A_s (\mathbf{k} \times \hat{\mathbf{s}}) +
A_p  (\mathbf{k} \times \hat{\mathbf{p}} )\right] \exp[i  \mathbf{k}\cdot\mathbf{r}] \\
\omega \mu \mathbf{H}_r &= \left[B_s (\mathbf{k}_r \times\hat{\mathbf{s}} )+
B_p (\mathbf{k}_r \times\hat{\mathbf{p}}_r) \right]  \exp[ i\mathbf{k}_r\cdot\mathbf{r}]\\
\omega \mu \mathbf{H}_t &= C_o (\mathbf{k}_o \times\hat{\mathbf{o}}
)\exp[i  \mathbf{k}_o\cdot\mathbf{r}] +
C_e (\mathbf{k}_e \times\hat{\mathbf{e}}) \exp[i \mathbf{k}_e\cdot\mathbf{r}].
\end{align}
Enforcing continuity of the components of the electric and magnetic
fields tangential to the interface (i.e. $\mathbf{E}\cdot \hat{\mathbf{g}}^*$, $\mathbf{E}\cdot
\hat{\mathbf{h}}^*$, $\mathbf{H}\cdot \hat{\mathbf{g}}^*$ and  $\mathbf{H}\cdot
\hat{\mathbf{h}}^*$) yields four equations viz. 
\begin{align}
A_s + B_s &= C_o \hat{\mathbf{o}} \cdot \hat{\mathbf{h}}^* + C_e
\hat{\mathbf{e}} \cdot \hat{\mathbf{h}}^* \label{eq:uniaxialcont1}\\
k_x(A_s - B_s) &= C_o (\mathbf{k}_o \times \hat{\mathbf{o}})\cdot
\hat{\mathbf{g}}^* + C_e (\mathbf{k}_e \times \hat{\mathbf{e}})\cdot
\hat{\mathbf{g}}^* \\
k_x(A_p - B_p) &= C_o  \hat{\mathbf{o}} \cdot \hat{\mathbf{g}}^* + C_e
\hat{\mathbf{e}} \cdot \hat{\mathbf{g}}^*\\
-A_p - B_p &= C_o (\mathbf{k}_o \times \hat{\mathbf{o}})\cdot
\hat{\mathbf{h}}^* + C_e (\mathbf{k}_e \times \hat{\mathbf{e}})\cdot
\hat{\mathbf{h}}^* \label{eq:uniaxialcont4}
\end{align}
where $k_x = \mathbf{k}\cdot \hat{\mathbf{x}}$ is the $x$ component of
$\mathbf{k}$.
We can define generalised reflection and transmission coefficients
according to \cite{Yeh1979}
\begin{align}
B_s &= r_{ss} A_s +r_{ps} A_p \\
B_p &= r_{sp} A_s +r_{pp} A_p \\
C_o &= t_{so} A_s +t_{po} A_p \\
C_e &= t_{se} A_s +t_{pe} A_p 
\end{align}
which can be found by solving
Eqs.~\myeqref{eq:uniaxialcont1}--\myeqref{eq:uniaxialcont4} yielding 
\begin{align}
r_{ss} &=  \frac{M^o_- N^e_- - M^e_- N^o_-}{M^o_+ N^e_- - M_+^e
	N_-^o}&&,& r_{pp} &= \frac{M_+^o N^e_+ - M^e_+ N^o_+}{M^o_+ N^e_- -
	M_+^e N_-^o} \label{eq:Fresnel1}\\ 
r_{ps} &=   \frac{M^o_+ M^e_- - M^o_- M^e_+}{M^o_+ N^e_- - M_+^e
	N_-^o}&&,& r_{sp} &= \frac{N_-^oN^e_+ - N^o_+ N^e_-}{M^o_+ N^e_- - M_+^e N_-^o}\\ 
t_{so} &=  \frac{2N^e_-}{M^o_+ N^e_- - M_+^e N_-^o} &&,& t_{se}
&= \frac{-2N^o_-}{M^o_+ N^e_- - M_+^e N_-^o}\\ 
t_{po} &=  \frac{-2M^e_+}{M^o_+ N^e_- - M_+^e N_-^o} &&,& t_{pe}
&= \frac{2 M^o_+}{M^o_+ N^e_- - M_+^e N_-^o}
\end{align}
and
\begin{align}
{M}_{\pm}^o &=  \hat{\mathbf{o}} \cdot  \hat{\mathbf{h}}^* \pm
({\mathbf{k}}_o \times \hat{\mathbf{o}}) \cdot
\hat{\mathbf{g}}^* / k_x\\
{M}_{\pm}^e &= \hat{\mathbf{e}} \cdot \hat{\mathbf{h}}^* \pm  (
\mathbf{k}_e  \times \hat{\mathbf{g}}) \cdot \hat{\mathbf{g}}^* /k_x\\
{N}_{\pm}^o &= - (\mathbf{k}_o \times \hat{\mathbf{o}}  ) \cdot
\hat{\mathbf{h}}^* / k \pm k \,\hat{\mathbf{o}} \cdot  \hat{\mathbf{g}}^* /k_x\\
{N}_{\pm}^e &= - (\mathbf{k}_e \times \hat{\mathbf{e}}  ) \cdot
\hat{\mathbf{h}}^* / k \pm k \, \hat{\mathbf{e}} \cdot
\hat{\mathbf{g}}^* /k_x. \label{eq:NePM_fresnel}
\end{align}
We note that Eqs.~\myeqref{eq:Fresnel1}--\myeqref{eq:NePM_fresnel} are
equivalent to those presented in \cite{Gu1993a} for propagating waves,
however differ for evanescent waves since we normalise unit vectors
such that $\mathbf{u}\cdot \mathbf{u}^* = 1$ as opposed to
$\mathbf{u}\cdot \mathbf{u} = 1$. We elect to use this normalisation
convention since it maintains the physical definition of the Fresnel
coefficients as the amplitude ratio of the respective field component \cite{Norrman2011}.

\section{Breit-Wigner line shape with material absorption}\label{app:Lorentzian}
We start by considering the scattering amplitude of a given WGM which
near resonance takes the form
\begin{align}
\eta_{\mathbf{v}}(\omega) =  -\frac{ \gamma_{{\mathbf{v}},\txtpow{rad}}/2}{\gamma_{{\mathbf{v}},\txtpow{rad}}/2-i(\omega - \omega_{\mathbf{v}})} \triangleq -\frac{1}{1-i  \beta(\omega)}  \label{eq:betadef}
\end{align}
when absorption in the resonator is neglected. To account for the
effect of absorption we must consider the dependence of
$\beta(\omega)$ on the complex refractive index
$\bar{n} = n + i \kappa$. Following
\cite{Johnson1993} we perform a Taylor expansion of $\beta(\omega,\bar{n})$
around the resonance frequency $\omega_{\mathbf{v}}$ and the real part of
the refractive index $n$ whereby
\begin{align}
\beta(\omega) \approx (\omega-\omega_{\mathbf{v}}) \, \partial_\omega
\beta(\omega_{\mathbf{v}},n) + i \kappa \, \partial_{\bar{n}} \beta(\omega_{\mathbf{v}},n)  \label{eq:betaexp}
\end{align}
and where $\beta(\omega_{\mathbf{v}},n) = 0$ has been used. Substituting
\eqref{eq:betaexp} into \eqref{eq:betadef} yields 
\begin{align}
\eta_{\mathbf{v}}(\omega) \approx - \frac{1}{1 +  \kappa \, \partial_{\bar{n}}
	\beta(\omega_{\mathbf{v}},n)  -i  (\omega-\omega_{\mathbf{v}}) \, \partial_\omega
	\beta(\omega_{\mathbf{v}},n)  }. \label{eq:betaderiv1}
\end{align}
Noting $\partial_\omega \beta(\omega_{\mathbf{v}},n) =
2/\gamma_{{\mathbf{v}},\txtpow{rad}}$, we observe that \eqref{eq:betaderiv1}
also possesses a Breit-Wigner line shape, albeit with a modified
line width of $\gamma_{\txtpow{tot}} = \gamma_{{\mathbf{v}},\txtpow{rad}} (1 + \kappa
\, \partial_{\bar{n}} \beta(\omega_{\mathbf{v}},n) ) =  \gamma_{{\mathbf{v}},\txtpow{rad}}
+  \gamma_{{\mathbf{v}},\txtpow{abs}}$. We can hence deduce $\kappa
\, \partial_{\bar{n}} \beta(\omega_{\mathbf{v}},n) =  \gamma_{{\mathbf{v}},\txtpow{abs}}/
\gamma_{{\mathbf{v}},\txtpow{rad}}$ whereby \eqref{eq:betaderiv1} yields
\eqref{eq:Lorentz_approx2} of the main text.

\section{Anisotropic Bethe-Schwinger equation}\label{app:BSderiv}
The eigenmodes of an open system can be written in the form
$\overline{\mathbf{E}}_{\mathbf{v}}(\mathbf{r},t) = \mathbf{E}_{\mathbf{v}}(\mathbf{r})
\exp[-i\overline{\omega} t]$, where $\overline{\omega} = \omega - i \gamma/2$ is
the complex eigenfrequency which describes both the real resonance
frequency $\omega$ and the resonance lifetime $1/\gamma$. The
associated magnetic field $\overline{\mathbf{H}}_{\mathbf{v}}(\mathbf{r},t)$ can
be similarly defined. The
eigenmodes must satisfy Maxwell's equations, which allowing for
anisotropic media reduce to
\begin{align}
\nabla \times \mathbf{E}_{\mathbf{v}} = i \overline{\omega} \mathbf{B}_{\mathbf{v}} 
\quad \mbox{and} \quad
& \nabla \times \mathbf{H}_{\mathbf{v}} = - i \overline{\omega}     \mathbf{D}_{\mathbf{v}} \label{eq:curleqs}
\end{align}
where $\mathbf{D} = \tensor{\epsilon} \mathbf{E}$, $\mathbf{B} = \tensor{\mu}
\mathbf{H}$, and $\tensor{\epsilon} = \tensor{\epsilon}(\mathbf{r})$ and $\tensor{\mu} =
\tensor{\mu}(\mathbf{r})$ are the electric permittivity and magnetic
permeability tensors describing the cavity and its surroundings. Modification of the
dielectric environment results in a modified set of
eigenmodes $\overline{\mathbf{E}}'_{\mathbf{v}} (\mathbf{r},t)
=\mathbf{E}'_{\mathbf{v}} (\mathbf{r})\exp[-i\overline{\omega}' t]$ (and similarly
for $\overline{\mathbf{H}}'_{\mathbf{v}}$), which must also satisfy an analogous set
of equations to \eqref{eq:curleqs} with the replacements
$\mathbf{E}_{\mathbf{v}}\rightarrow \mathbf{E}'_{\mathbf{v}}$, $\mathbf{H}_{\mathbf{v}}\rightarrow
\mathbf{H}'_{\mathbf{v}}$, $\tensor{\epsilon}(\mathbf{r}) \rightarrow
\tensor{\epsilon}{}'(\mathbf{r})$ and $\tensor{\mu}(\mathbf{r}) \rightarrow
\tensor{\mu}{}'(\mathbf{r})$. Subtracting these sets of
equations yields
\begin{align}
\nabla \times \delta\mathbf{E}_{\mathbf{v}} &= i \overline{\omega}'
\mathbf{B}_{\mathbf{v}}' - i \overline{\omega} \mathbf{B}_{\mathbf{v}} \\
\nabla \times \delta \mathbf{H}_{\mathbf{v}} &= - i \overline{\omega}'     \mathbf{D}_{\mathbf{v}}' + i \overline{\omega}     \mathbf{D}_{\mathbf{v}} 
\end{align}
where $\delta \mathbf{E}_{\mathbf{v}} = \mathbf{E}'_{\mathbf{v}} -
\mathbf{E}_{\mathbf{v}}$ (and similarly for $\delta
\mathbf{H}_{\mathbf{v}}$). Now forming the inner
products $\mathbf{H}^*_{\mathbf{v}} \cdot \nabla \times\delta \mathbf{E}_{\mathbf{v}}$ and
$\mathbf{E}^*_{\mathbf{v}} \cdot \nabla\times 
\delta \mathbf{H}_{\mathbf{v}}$, using  the vector
identity 
\begin{align}
\mathbf{A}\cdot (\nabla \times \mathbf{B}) =
\mathbf{B}\cdot(\nabla \times \mathbf{A}) - \nabla \cdot
(\mathbf{A}\times\mathbf{B})
\end{align}
and subtracting the resulting
expressions we obtain
\begin{align}
&\hspace{-0.2cm} i \overline{ \omega}^* ( \delta\mathbf{E}_{\mathbf{v}}\cdot\mathbf{D}^*_{\mathbf{v}} + 
\delta\mathbf{H}_{\mathbf{v}}\cdot \mathbf{B}^*_{\mathbf{v}} ) - \nabla\cdot(\mathbf{H}^*_{\mathbf{v}}\times
\delta\mathbf{E}_{\mathbf{v}} - \mathbf{E}^*_{\mathbf{v}} \times  \delta\mathbf{H}_{\mathbf{v}}) \nonumber\\
&\quad\quad =i \overline{\omega}' (\mathbf{E}^*_{\mathbf{v}} \cdot \mathbf{D}'_{\mathbf{v}} + 
\mathbf{H}^*_{\mathbf{v}} \cdot \mathbf{B}'_{\mathbf{v}}) - i \overline{\omega} (\mathbf{E}^*_{\mathbf{v}} \cdot \mathbf{D}_{\mathbf{v}} + 
\mathbf{H}^*_{\mathbf{v}} \cdot \mathbf{B}_{\mathbf{v}}).
\end{align}
Integrating over a spherical volume of radius $R_V$ and use of Gauss's theorem gives
\begin{align}
&-i\oint_S (\delta\mathbf{E}_{\mathbf{v}} \times
\mathbf{H}^*_{\mathbf{v}} + \mathbf{E}_{\mathbf{v}}^*\times \delta\mathbf{H}_{\mathbf{v}} ) \cdot d\mathbf{S} \\
&= \int_V \Big[ \overline{\omega}'( \mathbf{E}^*_{\mathbf{v}}\cdot
\mathbf{D}'_{\mathbf{v}} + \mathbf{H}^*_{\mathbf{v}}\cdot
\mathbf{B}'_{\mathbf{v}}) - \overline{\omega} (\mathbf{E}^*_{\mathbf{v}} \cdot \mathbf{D}_{\mathbf{v}} + 
\mathbf{H}^*_{\mathbf{v}} \cdot \mathbf{B}_{\mathbf{v}}) \nonumber\\
&\quad\quad  
-   \overline{\omega}^* (  \delta\mathbf{E}_{\mathbf{v}}\cdot
\mathbf{D}^*_{\mathbf{v}} + \delta\mathbf{H}_{\mathbf{v}}\cdot
\mathbf{B}^*_{\mathbf{v}})  \Big]dV \label{eq:BetheSchwinger}
\end{align}
from which both the resonance
frequency shift $\delta \omega = \omega'-\omega$ and mode broadening
$\delta \gamma = \gamma' - \gamma$ can be extracted by rewriting
\eqref{eq:BetheSchwinger} in the form
\begin{align}
\delta \overline{\omega}  = \Big[&- (\delta \overline{\omega}+
\overline{\omega}) N^{(1)}_{\mathbf{v}} -i
N^{(2)}_{\mathbf{v}}\nonumber\\
&\quad+ \overline{\omega}^*
N^{(3)}_{\mathbf{v}} + \overline{\omega}
\left(N^{(4)}_{\mathbf{v}}-N_{\mathbf{v}}^{(0)}\right) \Big] \Big/ N_{\mathbf{v}}^{(0)},
\end{align}
where we have let $\tensor{\epsilon}{}' (\mathbf{r}) = \tensor{\epsilon} (\mathbf{r}) +
\delta\tensor{\epsilon} (\mathbf{r})$ (and similarly for $\mu'$) and have defined the integral terms
\begin{align}
N^{(0)}_{\mathbf{v}}&= {\int_V  
	\mathbf{E}^*_{\mathbf{v}} \tensor{\epsilon}\mathbf{E}' _{\mathbf{v}} +  
	\mathbf{H}^*_{\mathbf{v}}\tensor{\mu}\mathbf{H}_{\mathbf{v}}'\,dV}\label{eq:BSint0}\\
N^{(1)}_{\mathbf{v}}&= {\int_V                           
	\mathbf{E}^*_{\mathbf{v}} \delta \tensor{\epsilon}      \mathbf{E}'
	_{\mathbf{v}}  + \mathbf{H}^*_{\mathbf{v}}  \delta \tensor{\mu}     \mathbf{H}' _{\mathbf{v}} \,dV}\label{eq:BSint1}\\
N^{(2)}_{\mathbf{v}}&={\oint_S (\delta\mathbf{E}_{\mathbf{v}} \times
	\mathbf{H}^*_{\mathbf{v}} + \mathbf{E}_{\mathbf{v}}^*\times \delta\mathbf{H}_{\mathbf{v}} ) \cdot d\mathbf{S}} \label{eq:BSint2}\\
N^{(3)}_{\mathbf{v}} &={\int_V  
	\delta\mathbf{E}_{\mathbf{v}} \tensor{\epsilon}{}^* \mathbf{E}^*_{\mathbf{v}}+
	\delta\mathbf{H}_{\mathbf{v}}\tensor{\mu}{}^* \mathbf{H}_{\mathbf{v}}^*  \,dV}\label{eq:BSint3}\\
N^{(4)}_{\mathbf{v}} &={\int_V  
	\mathbf{E}_{\mathbf{v}}^* \tensor{\epsilon}\mathbf{E}_{\mathbf{v}}+
	\delta\mathbf{H}_{\mathbf{v}}^*\tensor{\mu} \mathbf{H}_{\mathbf{v}}  \,dV}.\label{eq:BSint4}
\end{align}
After application of Gauss's theorem we let $R_V\rightarrow \infty$, such that the surface integral in $N_{\mathbf{v}}^{(2)}$ is taken at infinity as discussed in the main text. 

Practical WGM
resonators are typically fabricated from materials for which the
imaginary part of the permittivity is small. Although absorption can
still play an important role in determining the line width of WGMs it
is nevertheless safe to assume that $\epsilon$ is real when evaluating
the integrals in Eqs.~\myeqref{eq:BSint0}--\myeqref{eq:BSint4}. Noting that the electric
permittivity tensor $\epsilon$ is also symmetric and assuming
$\overline{\omega}  \approx \overline{\omega}^*$, i.e. restricting
attention to high $Q$ resonances it then follows that $\overline{\omega}^*
N^{(3)}_{\mathbf{v}} + \overline{\omega}
\left[N^{(4)}_{\mathbf{v}}-N^{(0)}_{\mathbf{v}}\right] = 0$. Making the further
assumption that $\delta\omega \ll \omega$ and $\delta\gamma \ll
\omega$ \eqref{eq:BetheSchwinger} can be simplified to 
\begin{align}
\delta \overline{\omega}  = - \left[\overline{\omega}
N^{(1)}_{\mathbf{v}} +i N^{(2)}_{\mathbf{v}}\right]\Big/N^{(0)}_{\mathbf{v}}.
\end{align}

\bibliographystyle{osajnl}

\end{document}